\documentclass[lettersize,journal]{IEEEtran}
\usepackage{amsmath,amsfonts,amssymb}
\usepackage{algorithm}
\usepackage{algorithmic}
\usepackage{array}
\usepackage[caption=false,font=normalsize,labelfont=sf,textfont=sf]{subfig}
\usepackage{textcomp}
\usepackage{stfloats}
\usepackage{url}
\usepackage{multirow}
\usepackage{tabularx,booktabs} 
\usepackage{verbatim}
\usepackage{graphicx}
\usepackage{cite}
\usepackage{makecell}
\usepackage{url}
\usepackage{relsize} 
\usepackage{color}
\usepackage{graphicx}
\makeatletter
\let\NAT@parse\undefined
\makeatother
\usepackage{hyperref}  

\begin{document}

\title{Max-Min Throughput in RSMA-Aided SWIPT for Sustainable Internet of Underground Things: A Deep Reinforcement Learning Approach}

\author{Kaiqiang~Lin,~\IEEEmembership{Member,~IEEE} and 
        Mohamed-Slim Alouini,~\IEEEmembership{Fellow,~IEEE}  
        
\thanks{This work was supported in part by King Abdullah University of Science and Technology. (\textit{Corresponding author: Kaiqiang Lin})}  

\thanks{K. Lin and M.-S. Alouini are with the Division of Computer, Electrical and Mathematical Sciences and Engineering, King Abdullah University of Science and Technology, Thuwal 23955, Saudi Arabia. E-mail: kaiqiang.lin@kaust.edu.sa; slim.alouini@kaust.edu.sa.} 
}

\maketitle
\begin{abstract}
Unmanned aerial vehicle (UAV)-enabled wireless-powered underground sensor networks (WPUSNs) allow underground devices (UDs) to harvest radio-frequency energy from a UAV via wireless energy transfer (WET) techniques, enabling sustainable Internet of underground things (IoUT) applications. This paper integrates rate-splitting multiple access (RSMA)-aided simultaneous wireless information and power transfer (SWIPT) into UAV-enabled WPUSNs to deliver information and energy to UDs. Within this framework, we investigate a max-min fairness problem that jointly optimizes the UAV transmit power, power allocation and message splitting, power-splitting ratios, and time allocation, aiming to maximize the minimum uplink throughput across UDs subject to per-UD downlink throughput requirements and a UAV energy-consumption constraint. To address this non-convex problem under dynamic underground environments, we develop an off-policy deep reinforcement learning approach based on the primal-dual truncated quantile critics (PD-TQC), where a TQC-based reward critic evaluates the fairness objective, cost critics estimate constraint costs, and an actor learns the resource-control policy with adaptive dual-variable updates. Simulations using \textit{in-situ} soil volumetric water content measurements from a center-pivot irrigation farm demonstrate that PD-TQC achieves the highest annual average minimum uplink throughput compared with four benchmarks while satisfying the per-frame constraints in over $96\%$ of the evaluated frames, and that this throughput advantage is preserved under different Rician factors, numbers of UDs, and downlink throughput thresholds.
\end{abstract}


\section{Introduction}
\label{sec:intro}
\subsection{Background}
\IEEEPARstart{W}{ireless} underground sensor networks (WUSNs) enable real-time monitoring of subterranean resources through wirelessly connected underground devices (UDs), facilitating various Internet of underground things (IoUT) applications, such as underground pipeline leakage detection, smart agriculture, post-disaster rescue, and border patrol~\cite{WUSNReview, IoUTReview, LinSAGUIN}. Unlike conventional cabled systems with aboveground loggers, fully buried WUSNs preserve an unobstructed ground surface and conceal sensing and communication components, reducing their exposure to agricultural machinery, adverse weather, and unauthorized interference. Such concealed deployment is particularly valuable for critical underground infrastructure, such as buried pipelines, where exposing monitoring equipment raises security concerns~\cite{VuranWUSNsreview, LinTerraFi2026}. However, compared with terrestrial wireless sensor networks, UDs with limited battery capacity require more energy for reliable communication with aboveground receivers due to severe attenuation in harsh underground soils~\cite{LinAdaptive}. Moreover, it is very cumbersome and even impractical to regularly replace the batteries of UDs~\cite{VuranWUSNsreview}.

To enable sustainable IoUT operation, radio-frequency (RF) wireless energy transfer (WET) technology has been extended into WUSNs, giving rise to wireless-powered underground sensor networks (WPUSNs), which continuously recharge UDs~\cite{LiuWPUSNs, LiuMIMOWPUSNs}. Several studies have investigated various techniques to enhance WPUSN performance. Specifically, the pioneering work in~\cite{LiuWPUSNs} conceptualized a multi-user WPUSN system, where UDs harvest RF energy from an aboveground power station to support wireless information transmission (WIT) to an access point. Herein, a time allocation strategy was developed to maximize network throughput under diverse data traffic demands. To further enhance throughput in highly heterogeneous underground environments, the authors in~\cite{LiuMIMOWPUSNs} leveraged the multiple-input multiple-output technology and jointly optimized the time allocation and beamforming based on channel state information (CSI) estimation. The work in~\cite{LinBSWPUSNs} integrated backscatter communication into WPUSNs to enable emergency data delivery and improve resource utilization, where time allocation among WET, backscattering, and WIT was designed to maximize throughput while ensuring communication reliability. Building on this framework, the authors in~\cite{LinRIS} further introduced a reconfigurable intelligent surface (RIS) to overcome the blockage of direct links and jointly optimized time scheduling and RIS phase shifts to enhance network reliability and achievable rates. To avoid the overhead of accurate CSI acquisition in the above studies, CSI-free multiantenna WET was investigated in~\cite{LinCSIfree, RAB} to power a large number of nearby UDs considering practical power budgets and two deployment strategies. Moreover, to support large-scale underground charging, unmanned aerial vehicles (UAVs) were introduced in~\cite{LinUAV}, leading to UAV-enabled WPUSNs, where a UAV equipped with a hybrid communication and power transfer module recharges the monitoring area and collects data from all UDs. For this setup, a hybrid CSI-free WET framework and a novel time allocation strategy were proposed to reduce the UAV energy consumption while satisfying the throughput requirements of all UDs. 

Despite extensive research on WPUSNs, most existing studies have overlooked downlink communication, which is indispensable in IoUT for delivering control commands and configuration updates, thereby enabling reliable network management and sustainable operation~\cite{Undergroundfield, LinLoRaWUSNs}. This motivates a simultaneous wireless information and power transfer (SWIPT)-enabled downlink architecture, where a controller simultaneously provides WET and WIT to UDs via either time switching or power-splitting (PS) protocols~\cite{SWIPTReview}. In this work, we focus on the PS protocol, as it has been shown to provide a superior rate-energy trade-off~\cite{RSSWIPT2020, SWIPTRP}. Specifically, each UD splits the received signal into two branches for information decoding (ID) and energy harvesting (EH) with an adjustable PS ratio. Although conventional time-division multiple access (TDMA) offers low receiver complexity, its orthogonal downlink transmission requires dedicated time allocation for individual UDs, limiting the flexibility of utilizing the available communication-frame duration to balance downlink throughput, EH, and subsequent uplink transmission. To address this limitation, we employ rate-splitting multiple access (RSMA), which uses linearly precoded rate-splitting (RS) at the transmitter to divide user messages into common and private parts, and applies successive interference cancellation (SIC) at the receivers to sequentially decode the common and private streams~\cite{RSSWIPT2020}. Consequently, RSMA enables robust and efficient interference management and resource allocation compared with TDMA, space-division multiple access (SDMA), and power-domain non-orthogonal multiple access (NOMA) schemes~\cite{MaoRSMAsurvey, RSSWIPT2020}. To establish a more complete and sustainable IoUT framework, we build on these technologies by incorporating an RSMA-aided SWIPT (RS-SWIPT) downlink into the UAV-enabled WPUSN, where a fixed-hover UAV simultaneously delivers control information and RF energy to the UDs, followed by TDMA-based uplink data transmission.

\subsection{Related Work and Motivations}
Inspired by the benefits of RSMA-enabled SWIPT, there has been a growing interest in studying RS-SWIPT-aided downlink transmissions in terrestrial networks and proposing several heuristic or suboptimal resource-allocation schemes to better exploit these advantages~\cite{SWIPTRP, RSSWIPT2020, RSSWIPT2022, CRSSWIPT2022, RSSWIPT2024, RSSWIPT2025}. Specifically, early work in~\cite{RSSWIPT2020} considered the joint optimization of precoders and PS ratios for transmit-power minimization subject to users’ minimum-rate and EH requirements, and obtained locally optimal solutions via particle swarm optimization. Building on this line, an RIS-assisted RSMA-SWIPT architecture was investigated in~\cite{SWIPTRP}, where the transmit-power minimization was formulated as a bilevel optimization problem and addressed through a genetic-algorithm-based outer search combined with a semidefinite-relaxation inner solver. Robust and energy-efficient beamforming for RSMA-SWIPT was then examined in~\cite{RSSWIPT2022} by jointly optimizing beamformers and PS ratios using a dual-layer iterative procedure. Cooperative extensions were explored in~\cite{CRSSWIPT2022}, which introduced a full-duplex cooperative RS-SWIPT scheme and solved the resulting transmit-power minimization problem with target-rate guarantees via an iterative successive convex approximation framework. More recently, the authors in~\cite{RSSWIPT2024} proposed a SWIPT-enabled ad hoc underlay RSMA cellular system empowered by RIS, and developed low-complexity alternating-optimization algorithms for a max-min throughput design that jointly optimizes common-rate allocation, transmit beamformers, PS factors, and reflecting coefficients. Furthermore, the authors in~\cite{RSSWIPT2025} considered hardware impairments in an RIS-assisted RSMA-SWIPT system and designed an alternating-optimization algorithm to tackle the resulting non-convex sum-rate maximization problem.

The aforementioned studies have substantially advanced resource allocation for RSMA-based SWIPT systems. However, their optimization solutions generally require repeatedly solving highly coupled and non-convex resource-allocation problems as the channel and system conditions change. This computational burden becomes particularly challenging in UAV-assisted WPUSNs, where wireless channel conditions vary with small-scale fading and are further affected by time-varying underground soil properties, such as volumetric water content (VWC) variations induced by precipitation and irrigation~\cite{ZhaoMADRL, LinDtSDRL}. These characteristics motivate the adoption of learning-based resource-control approaches, which can learn effective control policies directly from interactions with the environment and provide computationally efficient decisions under stochastic and time-varying channel conditions.

In this context, deep reinforcement learning (DRL) has recently emerged as a promising paradigm for intelligent resource optimization in RSMA, SWIPT, and RSMA-based SWIPT networks~\cite{GreedyApp, RSMADQNPPO, UAVSWIPTSAC, RISSWIPTSAC, RSSWIPTDQ2025, RSSWIPTPPO2023}. For instance, in~\cite{GreedyApp, RSMADQNPPO}, the achievable sum rate of RSMA-enabled networks was maximized by jointly optimizing power allocation and rate control using a DRL framework based on the proximal policy optimization (PPO) algorithm. More recently, soft actor-critic (SAC)-based DRL frameworks were investigated for SWIPT systems to learn resource-control policies, targeting either system-level energy consumption reduction~\cite{UAVSWIPTSAC} or achievable rate maximization~\cite{RISSWIPTSAC}. For RS-SWIPT-aided downlink transmission, Q-learning was adopted in~\cite{RSSWIPTDQ2025} to jointly optimize uplink/downlink scheduling, transmit power, PS ratios, and uplink decoding order to maximize packet deliveries, while the authors in~\cite{RSSWIPTPPO2023} utilized the PPO algorithm in RIS-assisted RS-SWIPT networks to maximize energy efficiency under transmit power and quality-of-service constraints, demonstrating significant gains over conventional schemes. However, these studies focus exclusively on terrestrial networks, leaving a research gap in DRL-based joint resource optimization for our considered UAV-enabled WPUSNs with RS-SWIPT-aided downlink transmissions.

\subsection{Contributions}
To enable sustainable IoUT operation, we extend the UAV-enabled WPUSN framework in~\cite{LinUAV} by incorporating an RS-SWIPT-aided downlink. Specifically, a fixed-hover UAV performs RSMA-based SWIPT to simultaneously deliver control information and RF energy to UDs, while each UD employs an adjustable PS ratio to divide the received signal between ID and EH and subsequently uses the harvested energy for TDMA-based uplink WIT. This architecture introduces the challenge of jointly optimizing power allocation, message splitting, time-slot scheduling, and PS control under time-varying soil-air channels to enhance system performance. Motivated by the capability of DRL, we propose to exploit a DRL-based framework to address this non-convex joint optimization problem. To the best of our knowledge, this is the first work to investigate DRL-driven joint resource optimization for UAV-enabled WPUSNs with RS-SWIPT-aided downlink transmission. The main contributions are summarized as follows:
\begin{enumerate}
    \item We conceptualize a UAV-enabled WPUSN system that integrates an RS-SWIPT-aided downlink with a TDMA-based uplink to support sustainable IoUT operation. Under a practical nonlinear EH model, we formulate a max-min throughput optimization problem that maximizes the minimum uplink throughput among all UDs by jointly optimizing the UAV transmit power, power allocation and message splitting for RSMA, power-splitting ratios at SWIPT receivers, and downlink/uplink time allocation under prescribed downlink-throughput and UAV-energy constraints.

    \item To address the stochastic resource-allocation problem under time-varying soil-air channels, we develop an off-policy DRL framework based on primal-dual truncated quantile critics (PD-TQC). Specifically, PD-TQC employs an ensemble of TQC-based reward critics to evaluate the max-min throughput objective, utilizes two pairs of twin cost critics to estimate the corresponding discounted constraint-cost returns, and trains a stochastic actor to learn the resource-control policy with adaptive dual-variable updates.
    
    \item Through trace-driven simulations using \textit{in-situ} soil VWC measurements from a center-pivot irrigation farm~\cite{VWCdata} for training and out-of-sample evaluation, we demonstrate that the proposed PD-TQC algorithm consistently delivers higher annual average minimum uplink throughput than four well-established benchmark schemes, while satisfying the per-UD downlink throughput and UAV residual-energy requirements in $96.99\%$ and $99.72\%$ of the evaluated frames, respectively. Furthermore, the performance gain is preserved across different Rician factors, numbers of UDs, and downlink throughput targets in the considered scenario.
\end{enumerate}

\subsection{Organization and Notations}
The remainder of this article is organized as follows. Section~\ref{sec:sysmod} presents the system model and formulates the optimization problem. Section~\ref{sec:mdp} casts the problem as a Markov decision process (MDP), and Section~\ref{sec:framework} develops a PD-TQC-based DRL approach for its solution. Finally, Section~\ref{sec:res} analyzes the simulation results, and Section~\ref{sec:conclusion} concludes the article.

\textit{Notation:} Boldface lowercase and uppercase letters are used for vectors and matrices, respectively, such as $\mathbf{x}$ and $\mathbf{X}$. $\mathbb{R}$ and $\mathbb{C}$ refer to the sets of real and complex numbers, respectively. Superscripts $(\cdot)^H$ and $(\cdot)^T$ indicate the conjugate transpose and transpose operations. The operators $\min(\cdot)$ and $\max(\cdot)$ return the minimum and maximum values, respectively. $\mathbb{E}[\cdot]$ represents statistical expectation, and the positive-part operator is defined as $[\cdot]_+ \triangleq \max(\cdot,0)$. $\|\cdot\|$ stands for the Euclidean norm, while $|\cdot|$ gives the absolute value. $\mathcal{CN}(\mu,\sigma^2)$ designates a circularly symmetric complex Gaussian distribution with mean $\mu$ and variance $\sigma^2$. The projection onto an interval $[a,b]$ is written as $\Pi_{[a,b]}(\cdot)$, and $\bigcup$ denotes concatenation across the ensemble. Finally, $\mathrm{Sort}(\cdot)$, $\mathrm{Trunc}_{N}(\cdot)$, and $\mathrm{Mean}(\cdot)$ correspond to the sorting, truncation (i.e., discarding the largest $N$ elements), and averaging operations, respectively.

\section{System Model}
\label{sec:sysmod}
\begin{figure}[!t]
\centering
\includegraphics[width=2.4in]{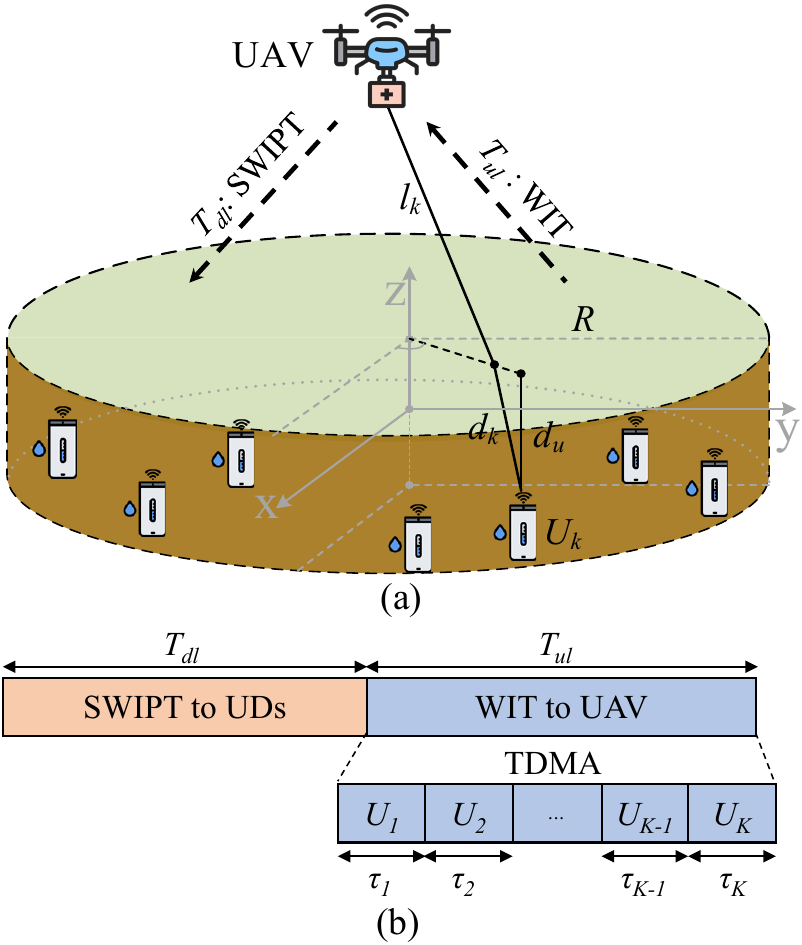}
\caption{A UAV-enabled WPUSN system integrating an RS-SWIPT-aided downlink and a TDMA-based uplink. (a) System model. (b) Time block.}
\label{fig_sysarch}
\end{figure}

\begin{figure}[!t]
\centering
\includegraphics[width=3.45in]{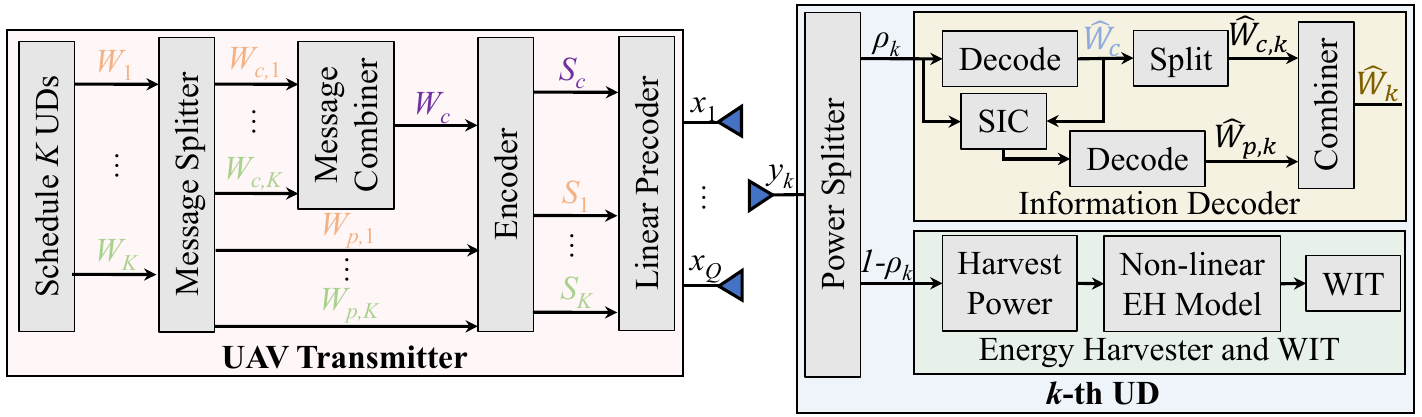}
\caption{RS-SWIPT-aided downlink transmission architecture based on one-layer RSMA.}
\label{fig_swiptarch}
\end{figure}

As depicted in Fig.~\ref{fig_sysarch}(a), we consider a multi-user UAV-enabled WPUSN consisting of a $Q$-antenna UAV and $K$ single-antenna UDs, indexed by $\mathcal{K}=\{1,2,\ldots,K\}$. The UAV operates at a fixed altitude $H$ and hovers above the center of the monitoring area during the considered communication service interval.\footnote{This fixed-hover configuration allows us to focus on joint communication-and-energy resource allocation, while the promising extension to UAV trajectory optimization is left for future work.}  The UDs are distributed within a circular area of radius $R$ and are buried at the same depth $d_u$. The UAV is equipped with a hybrid information-and-power transmission module, while each UD employs a PS receiver for simultaneous ID and EH. Under the adopted one-layer RSMA architecture, each UD decodes the common stream before recovering its intended private stream through a single SIC operation~\cite{MaoRSMAsurvey}. Moreover, SIC-free RSMA can be realized through appropriate receiver and precoder designs~\cite{SICfree}, while harvested RF energy can support receiver operation, facilitating the application of RSMA to energy-constrained UDs.

As illustrated in Fig.~\ref{fig_sysarch}(b), each communication frame has duration $T=T_{dl}+T_{ul}$. During the downlink phase of duration $T_{dl}$, the fixed-hover UAV employs RSMA-based SWIPT to simultaneously deliver information and RF energy to all UDs. The UDs then use the harvested energy for uplink WIT during the interval $T_{ul}$ using TDMA. Specifically, the $k$-th UD transmits its sensed data during an allocated time interval $\tau_k$, where $\sum_{k=1}^{K}\tau_k=T_{ul}$, thereby avoiding inter-user interference in the uplink. Finally, the UAV forwards the collected data to the network server via high-altitude platforms or satellites. In this study, we assume that perfect CSI of the downlink and uplink channels is available at the UAV at the beginning of each frame for downlink precoding, uplink combining, and resource allocation.\footnote{Although CSI acquisition is challenging due to the limited energy of UDs and the severe soil attenuation~\cite{LinCSIfree}, the energy for pilot transmission can be replenished through WET, while the estimation overhead can be reduced by exploiting the slowly varying soil-dependent attenuation and the quasi-static channel component determined by the fixed UAV and UD positions. Efficient CSI acquisition strategies, such as energy-aware pilot allocation and channel prediction based on historical CSI and VWC records, can further improve the estimation accuracy and reliability, which is left for future work.}

\subsection{RS-SWIPT-Aided Downlink Transmission}
During the downlink phase, the UAV adopts a one-layer RSMA strategy, as illustrated in Fig.~\ref{fig_swiptarch}, to simultaneously deliver information and energy to all $K$ UDs. Specifically, the message $W_k$ intended for the $k$-th UD is split into a common part $W_{c,k}$ and a private part $W_{p,k}$. All common parts $\{W_{c,1}, \ldots, W_{c,K}\}$ are combined into a super-common message $W_c$ and encoded into a common stream $s_c$ using a common codebook, whereas each private part $W_{p,k}$ is independently encoded into a private stream $s_k$, with $\mathbb{E}[|s_c|^2]=\mathbb{E}[|s_k|^2]=1$. This message-splitting and encoding structure enables each UD to first decode $s_c$ and then decode its intended private stream $s_k$ via SIC.

Let $\mathbf{P}=[\mathbf{p}_c,\mathbf{p}_1,\ldots,\mathbf{p}_K]\in\mathbb{C}^{Q\times (K+1)}$ denote the linear precoding matrix at the UAV, where $\mathbf{p}_c$ and $\mathbf{p}_k$ represent the beamforming vectors for the common stream and the $k$-th private stream, respectively, and $\|\mathbf{p}_c\|^2=\|\mathbf{p}_k\|^2=1$. Accordingly, the downlink transmit signal is given by
\begin{equation}
\mathbf{x}=\sqrt{\beta_c P_t}\,\mathbf{p}_c s_c+\sum_{k=1}^{K}\sqrt{\beta_k P_t}\,\mathbf{p}_k s_k, \label{EqTxSignal}
\end{equation}
where $P_t$ is the total UAV transmit power, and $\beta_c\ge0$ and $\beta_k\ge0$ are the power-allocation coefficients satisfying $\beta_c+\sum_{k=1}^{K}\beta_k=1$. Following a low-complexity design principle, we adopt maximum-ratio transmission (MRT) to determine the beam directions. Specifically, the common-stream beamformer is aligned with the superposition of all UDs' channels to facilitate reliable decoding at every UD, while each private-stream beamformer is aligned with its intended UD's channel. Accordingly, the RF signal received at the input of the PS receiver of the $k$-th UD is modeled as
\begin{equation}
y_k^{\mathrm{RF}}=\mathbf{h}_k^{H}\mathbf{x},
\end{equation}
where $\mathbf{h}_k\in\mathbb{C}^{Q\times 1}$ denotes the downlink channel from the UAV to the $k$-th UD. Following the commonly used PS receiver model in~\cite{LiuPSSWIPT}, we assume that the pre-splitting antenna noise is negligible relative to the post-splitting processing noise. Accordingly, the equivalent baseband signal at the ID branch is
\begin{equation}
y_k^{\mathrm{ID}} = \sqrt{\rho_k}\mathbf{h}_k^{H}\mathbf{x} + n_k, \label{IDsignal}
\end{equation}
where $\rho_k\in[0,1]$ is the PS ratio and $n_k\sim\mathcal{CN}(0,\sigma^2)$ denotes the equivalent post-splitting ID processing noise.

According to the validated air-soil propagation models in~\cite{Undergroundfield, LinBSWPUSNs}, the UAV-to-UD channel can be modeled as
\begin{equation}
    \mathbf{h}_k = \boldsymbol{\delta}_k^{dl} \sqrt{\frac{G_{uav} G_k}{L_{k}^{r-dl}L_{k}^{u} L_k^{a}}},
\end{equation}
where $\boldsymbol{\delta}_k^{dl}$ denotes the small-scale fading vector from the UAV to the $k$-th UD. It follows a Rician model given by
\begin{equation}
\boldsymbol{\delta}_k^{dl} = \sqrt{\frac{\kappa_R}{\kappa_R+1}} \bar{\boldsymbol{\delta}}_k^{dl} + \sqrt{\frac{1}{\kappa_R+1}}
\mathbf{g}_k^{dl},
\end{equation}
where $\kappa_R$ is the Rician factor, $\bar{\boldsymbol{\delta}}_k^{dl}$ denotes the deterministic line-of-sight (LoS) component, and $\mathbf{g}_k^{dl}\sim\mathcal{CN}(\mathbf{0},\mathbf{I}_Q)$ denotes the non-LoS component. Moreover, $G_{uav}$ and $G_k$ denote the antenna gains of the UAV and the $k$-th UD, respectively, while $L_{k}^{r-dl}$, $L_{k}^{u}$, and $L_{k}^{a}$ represent the air-soil refraction loss, underground propagation loss, and air propagation loss, respectively. Specifically,
\begin{align}
    L_{k}^{r-dl} &= \left(\frac{\sqrt{\left(\sqrt{\varepsilon'^{2}+\varepsilon''^{2}}+\varepsilon' \right)/2}+1}{4}\right)^2, \\
    L_{k}^{u} &= \left(2 \kappa d_k\right)^{2} e^{2 \nu d_k}, \label{EqUPL}\\
    L_k^{a} &= \left(\frac{4 \pi f l_k}{c}\right)^\varsigma,
\end{align}
where $d_k$ and $l_k$ denote the underground and air propagation distances from the UAV to the $k$-th UD, respectively, $f$ is the carrier frequency, $c$ is the speed of light, and $\varsigma$ denotes the air path-loss exponent. The parameters
$\nu$ and $\kappa$ denote the attenuation and phase constants of electromagnetic propagation in soil, respectively, and are given by 
\begin{align}
    \nu &= 2 \pi f \sqrt{\frac{\mu_{r} \mu_{0} \varepsilon' \varepsilon_{0}} {2}\left[\sqrt{1+\left(\varepsilon''/\varepsilon'\right)^{2}}-1 \right]}, \\
    \kappa &= 2 \pi f \sqrt{\frac{\mu_{r} \mu_{0} \varepsilon' \varepsilon_{0}} {2}\left[\sqrt{1+\left(\varepsilon''/\varepsilon'\right)^{2}}+1 \right]},
\end{align}
where $\mu_{r}$ is the soil relative permeability, $\mu_{0}$ is the free-space permeability, $\varepsilon_{0}$ is the free-space permittivity, while $\varepsilon'$ and $\varepsilon''$ are the real and imaginary parts of the soil relative permittivity, respectively, i.e., $\varepsilon=\varepsilon'+j\varepsilon''$. Note that $\varepsilon$ can be computed using the mineralogy-based soil dielectric model proposed in~\cite{MBSDM}. In particular, the model requires only three input parameters: the VWC, the RF operating frequency, and the clay percentage.

As illustrated in Fig.~\ref{fig_swiptarch}, each UD employs a PS receiver, where a fraction $\rho_k\in[0,1]$ of the received RF power is routed to the ID branch, while the remaining fraction $1-\rho_k$ is used for EH. Under the one-layer RSMA architecture, the $k$-th UD first decodes the common stream by treating all private streams as interference and then applies SIC to remove the decoded common stream before decoding its intended private stream. Based on the ID signal model in Eq.~\eqref{IDsignal}, the achievable downlink rates of the common and private streams at the $k$-th UD are respectively given by
\begin{align}
R_{c,k}^{dl} &= B\frac{T_{dl}}{T}\log_2\!\left(1+\frac{\rho_k \beta_c P_t|\mathbf{h}_k^H\mathbf{p}_c|^2}{\sum_{j=1}^{K}\rho_k \beta_j P_t|\mathbf{h}_k^H\mathbf{p}_j|^2+\sigma^2}\right), \label{EqRScomrate}\\
R_{p,k}^{dl} &= B\frac{T_{dl}}{T}\log_2\!\left(1+\frac{\rho_k \beta_k P_t|\mathbf{h}_k^H\mathbf{p}_k|^2}{\sum_{j\neq k}\rho_k \beta_j P_t|\mathbf{h}_k^H\mathbf{p}_j|^2+\sigma^2}\right). \label{EqRSprirate}
\end{align}
Herein, $B$ denotes the system bandwidth, while $T_{dl}/T$ represents the fraction of each communication frame allocated to downlink transmission. Hence, $R_{c,k}^{dl}$ and $R_{p,k}^{dl}$ are expressed in bit/s. Let $\alpha_k$ denote the fraction of the common-stream rate allocated to the $k$-th UD, where $\alpha_k\ge 0$ and $\sum_{k=1}^{K}\alpha_k=1$. Since the common stream must be decoded by all UDs, its achievable rate is limited by the minimum common-stream decoding rate among the UDs, i.e., $R_c^{dl}=\min_{k\in\mathcal{K}}R_{c,k}^{dl}$. Consequently, the effective downlink rate of the $k$-th UD is
\begin{equation}
R_k^{dl}=\alpha_k R_c^{dl}+R_{p,k}^{dl}.
\label{EqRSsumrate}
\end{equation}
Meanwhile, the received RF power at the input of the PS receiver of the $k$-th UD is given by
\begin{equation}
P_k^{rec}=P_t\!\left(\beta_c |\mathbf{h}_k^H\mathbf{p}_c|^2+\sum_{j=1}^{K}\beta_j |\mathbf{h}_k^H\mathbf{p}_j|^2\right).
\end{equation}
Following the practical nonlinear EH model in~\cite{PracticalEH}, the energy harvested during the downlink phase is given by
\begin{equation}
E_k = T_{dl}\,\frac{\frac{M}{1+e^{-a\left((1-\rho_k) P_k^{rec}-b\right)}}-\frac{M}{1+e^{ab}}}
{1-\frac{1}{1+e^{ab}}}, \label{EqEHModel}
\end{equation}
where $M$ denotes the saturation power, while $a$ and $b$ are circuit-dependent constants.

\subsection{TDMA-Based Uplink Transmission}
In the uplink phase, each UD transmits its sensed data to the UAV using the harvested energy via TDMA. Let $\tau_k$ denote the uplink time allocated to the $k$-th UD, satisfying $\sum_{k=1}^{K}\tau_k=T_{ul}$. The $k$-th UD uses a fixed fraction $\varphi_k\in[0,1]$ of its harvested energy $E_k$ for uplink RF transmission, while the remaining energy accounts for circuit operation. Accordingly, the uplink transmit power of the $k$-th UD is given by
\begin{equation}
p_k^{ul}=\frac{\varphi_k E_k}{\tau_k}.
\end{equation}
Let $\mathbf{h}_k^{ul}\in\mathbb{C}^{Q\times 1}$ denote the uplink channel from the $k$-th UD to the UAV, which is modeled as a Rician fading channel incorporated with the large-scale attenuation, i.e.,
\begin{equation}
\mathbf{h}_k^{ul} = \boldsymbol{\delta}_k^{ul} \sqrt{\frac{G_kG_{uav}} {L_k^{r-ul}L_k^{u}L_k^{a}}},
\end{equation}
where $\boldsymbol{\delta}_k^{ul}$ denotes the small-scale Rician fading vector and is assumed to have the same Rician factor $\kappa_R$ as in the downlink. According to the soil-to-air propagation model adopted in~\cite{Undergroundfield, LinBSWPUSNs}, we set $L_k^{r-ul}=1$ for the soil-air interface in the uplink.

Assuming that the UAV applies maximum-ratio combining (MRC) with $\mathbf{w}_k=\mathbf{h}_k^{ul}/\|\mathbf{h}_k^{ul}\|$, the achievable uplink rate of the $k$-th UD is given by
\begin{equation}
R_k^{ul} = B\frac{\tau_k}{T} \log_2\!\left(1+\frac{p_k^{ul}\|\mathbf{h}_k^{ul}\|^2}{\sigma^2}\right).
\label{EquplinkT}
\end{equation}

\subsection{UAV Energy Consumption Model}
We consider a quasi-static UAV operation within each communication frame, where the UAV hovers above the center of the monitoring area to (i) transmit the RS-SWIPT signal during the downlink interval $T_{dl}$ and (ii) receive the TDMA-based uplink transmissions during the uplink interval $T_{ul}$. Following the rotary-wing UAV energy consumption model in~\cite{UAVEnergyModel}, the frame-level UAV energy consumption is modeled as
\begin{equation}
E_{uav}=P_{hov}T+P_tT_{dl},
\label{EqUAVE}
\end{equation}
where $P_{hov}=P_0+P_i$ denotes the hovering power, $P_0$ and $P_i$ are constants determined by the UAV platform and operating conditions~\cite{UAVEnergyModel}, while $P_tT_{dl}$ accounts for the RF transmit energy during the downlink phase. Other communication-related circuit energy, such as signal-processing and baseband-circuit consumption, is typically much smaller than the hovering energy and is therefore neglected in this study~\cite{UAVEignore}. Accordingly, $E_{uav}$ characterizes the per-frame UAV energy expenditure under the considered quasi-static hovering operation and is used in the subsequent resource-allocation formulation.

\subsection{Problem Formulation}
Our objective is to maximize the minimum uplink throughput among all UDs under downlink-throughput and UAV-energy requirements. Let $E_{bat}$ denote the UAV
energy budget available for each considered communication frame. With the optimization variables $\{\boldsymbol{\alpha}, \boldsymbol{\beta}, \boldsymbol{\tau}, \boldsymbol{\rho}, P_t, T_{dl}\}$, where $\boldsymbol{\alpha}=[\alpha_1,\dots,\alpha_K]$ is the RSMA common-rate allocation vector, $\boldsymbol{\beta}=[\beta_c,\beta_1,\dots,\beta_K]$ denotes the downlink transmit-power allocation ratios, $\boldsymbol{\tau}=[\tau_1,\dots,\tau_K]$ is the uplink TDMA time-allocation vector, and $\boldsymbol{\rho}=[\rho_1,\dots,\rho_K]$ collects the PS ratios at the UDs for balancing ID and EH during the downlink, the optimization problem can be formulated as
\begin{subequations}
\begin{align}
\max_{\boldsymbol{\alpha},\boldsymbol{\beta},\boldsymbol{\tau},
\boldsymbol{\rho},P_t,T_{dl}}
&\quad \min_{k\in\mathcal{K}} R_k^{ul}
\label{obj_main}\\
\text{s.t.}\quad
& R_k^{dl}\ge R_{\min},\quad \forall k,
\label{c0}\\
& \beta_c+\sum_{k=1}^{K}\beta_k=1,\quad
0\le\beta_c,\beta_k\le1,
\label{c2}\\
& 0\le P_t\le P_{\max},
\label{c3}\\
& \sum_{k=1}^{K}\tau_k+T_{dl}=T,\quad
T_{dl}\ge0,\ \tau_k>0,\ \forall k,
\label{c4}\\
& 0\le\rho_k\le1,\quad \forall k,
\label{c5}\\
& \sum_{k=1}^{K}\alpha_k=1,\quad
0\le\alpha_k\le1,\quad \forall k,
\label{c6}\\
& E_{bat}-E_{uav}\ge E_{th},
\label{c7}\\
& p_k^{ul}\le P_{k,\max},\quad \forall k.
\label{c8}
\end{align}
\end{subequations}
Constraint~\eqref{c0} specifies the minimum downlink rate requirement for each UD. Constraint~\eqref{c2} enforces the normalization and feasibility of the RSMA power-allocation ratios. Constraint~\eqref{c3} limits the UAV transmit power. Constraint~\eqref{c4} ensures that the downlink and uplink transmission intervals jointly occupy the communication frame of duration $T$. Constraint~\eqref{c5} bounds the PS ratios at the UDs, while constraint~\eqref{c6} normalizes the common-rate allocation factors. Constraint~\eqref{c7} preserves a prescribed residual UAV energy margin $E_{th}$ within the considered frame-level energy budget. Finally, constraint~\eqref{c8} imposes a per-UD maximum uplink transmit power constraint considering practical transceiver hardware limitations and regional regulatory restrictions.
 
\section{MDP Formulation}
\label{sec:mdp}
The max-min resource allocation problem in \eqref{obj_main}--\eqref{c8} is highly coupled and non-convex, since the performance of the RS-SWIPT-aided downlink and the TDMA-based uplink is jointly determined by the time allocation, the UAV transmit-power control, the power-splitting ratios at the UDs, and the RSMA rate/power partitioning variables. Moreover, the wireless channels and soil conditions (i.e., VWC) vary over time, thereby introducing stochastic variations into the system. Inspired by recent DRL-enabled designs for RS-SWIPT-aided networks, we formulate the resulting sequential decision-making task as an MDP defined by the tuple $(\mathcal{S},\mathcal{A},\mathcal{P},\mathcal{R},\gamma)$, where $\mathcal{S}$ and $\mathcal{A}$ denote the state and action spaces, $\mathcal{P}$ is the unknown state transition kernel, $\mathcal{R}$ is the reward function, and $\gamma\in[0, 1)$ is the discount factor. In this framework, the UAV acts as an agent that interacts with the environment and learns a resource control policy without requiring explicit knowledge of $\mathcal{P}$.

\subsection{State Space}
In our implementation, each decision epoch is indexed by one daily VWC sample from the soil database. At epoch $t$, the daily-mean VWC is used to update the soil dielectric parameters and the resulting underground attenuation~\cite{VWCdata}, while the selected resource-control action is evaluated over the corresponding communication frame. Let $t\in\{0,1,2,\ldots\}$ index the resulting trace-driven decision epochs. At epoch $t$, the UAV observes a state vector $\mathbf{s}_t$ and selects a continuous control action $\mathbf{a}_t$, and then the environment evaluates the RS-SWIPT-aided downlink and TDMA-based uplink performance and transitions to $\mathbf{s}_{t+1}$. The state $\mathbf{s}_t\in\mathcal{S}$ is defined by
\begin{equation}
\mathbf{s}_t=\left[\mathbf{g}^{c}_t,\mathbf{g}^{p}_t,\mathbf{g}^{ul}_t,\boldsymbol{\iota}^{p}_t,r_{t-1},\bar{E}_t,\upsilon_t\right],
\label{eq:state_def_new}
\end{equation}
where (i) $\mathbf{g}^{c}_t=[g^{c}_{1,t},\ldots,g^{c}_{K,t}]$, with $g^{c}_{k,t}\triangleq|\mathbf{h}_{k,t}^{H}\mathbf{p}_{c,t}|^2$, represents the effective downlink gains for common-stream decoding; (ii) $\mathbf{g}^{p}_t=[g^{p}_{1,t},\ldots,g^{p}_{K,t}]$, with $g^{p}_{k,t}\triangleq|\mathbf{h}_{k,t}^{H}\mathbf{p}_{k,t}|^2$, characterizes the desired private-stream gains under MRT precoding; (iii) $\mathbf{g}^{ul}_t=[g^{ul}_{1,t},\ldots,g^{ul}_{K,t}]$, with $g^{ul}_{k,t}\triangleq\|\mathbf{h}^{ul}_{k,t}\|^2$, captures the effective uplink gains under MRC; (iv) $\boldsymbol{\iota}^{p}_t=[\iota^{p}_{1,t},\ldots,\iota^{p}_{K,t}]$ describes the normalized aggregate private-stream interference, where $\iota^{p}_{k,t}\triangleq\frac{\sum_{j\neq k}|\mathbf{h}_{k,t}^{H}\mathbf{p}_{j,t}|^2}{|\mathbf{h}_{k,t}^{H}\mathbf{p}_{k,t}|^2}$; (v) $r_{t-1}$ is the reward obtained at the preceding decision epoch; (vi) $\bar{E}_t$ is the normalized residual-energy indicator from the preceding communication frame, defined as $\bar{E}_t\triangleq(E_{bat}-E_{uav,t-1})/E_{bat}$ for $t\ge1$, with $\bar{E}_0=1$; and (vii) $\upsilon_t$ denotes the daily-mean VWC. The state variables $\mathbf{g}_t^c$, $\mathbf{g}_t^p$, and $\boldsymbol{\iota}_t^p$ jointly characterize the principal signal and interference conditions governing the downlink signal-to-interference-plus-noise ratio without explicitly including all individual private-stream interference gains. Together with $\mathbf{g}_t^{ul}$, $r_{t-1}$, $\bar{E}_t$, and $\upsilon_t$, these variables provide relevant information for joint resource allocation while maintaining a manageable state dimension for policy learning.

\subsection{Action Space}
\label{subsec:action}
At each decision epoch $t$, the UAV selects continuous control variables to configure the RS-SWIPT-aided downlink and TDMA-based uplink. The action $\mathbf{a}_t\in\mathcal{A}$ is defined as
\begin{equation}
\mathbf{a}_t=\left[\boldsymbol{\beta}_t,\boldsymbol{\alpha}_t,\boldsymbol{\rho}_t,\boldsymbol{\tau}_t,\theta_t,\eta_t\right],
\label{eq:action_def_new}
\end{equation}
where $\boldsymbol{\beta}_t=[\beta_{c,t},\beta_{1,t},\ldots,\beta_{K,t}]$ is the RSMA power partition satisfying $\beta_{c,t}+\sum_{k=1}^{K}\beta_{k,t}=1$, $\boldsymbol{\alpha}_t=[\alpha_{1,t},\ldots,\alpha_{K,t}]$ is the common-rate allocation satisfying $\sum_{k=1}^{K}\alpha_{k,t}=1$, and $\boldsymbol{\rho}_t=[\rho_{1,t},\ldots,\rho_{K,t}]$ collects the PS ratios. Moreover, $\theta_t\in(0,1)$ denotes the normalized downlink-time fraction such that $T_{dl,t}=T\theta_t$, while $\boldsymbol{\tau}_t=[\tau_{1,t},\ldots,\tau_{K,t}]$ denotes the uplink TDMA time allocation satisfying $\sum_{k=1}^{K}\tau_{k,t}=T(1-\theta_t)$ and $\tau_{k,t}>0$. Finally, $\eta_t\in[0,1]$ is the normalized UAV transmit-power ratio such that $P_t=\eta_tP_{\max}$.

To satisfy the simplex and box constraints in~\eqref{c2}--\eqref{c6} by construction, the actor parameterizes the action components using Dirichlet and Beta distributions. Specifically, $\boldsymbol{\beta}_t$ and $\boldsymbol{\alpha}_t$ are sampled from Dirichlet distributions, while $\boldsymbol{\rho}_t$, $\theta_t$, and $\eta_t$ are sampled from Beta distributions. For uplink time allocation, the actor generates a simplex-valued vector and rescales it by $T(1-\theta_t)$ to obtain $\boldsymbol{\tau}_t$. The uplink peak-power constraint~\eqref{c8} is enforced by
\begin{equation}
p_{k,t}^{ul}=\min\!\left(\frac{\varphi_kE_{k,t}}{\tau_{k,t}},P_{k,\max}\right),\quad \forall k\in\mathcal{K},
\label{eq:ul_clip}
\end{equation}
which directly ensures $p_{k,t}^{ul}\le P_{k,\max}$. When the uplink transmit-power limit is reached, additional harvested energy no longer improves the uplink throughput for a fixed time allocation, thereby discouraging unnecessary EH and favoring resource allocation toward uplink TDMA transmission, subject to the downlink throughput and UAV residual-energy constraints.

\subsection{Reward and Constraint Costs}
\label{subsec:reward}
The objective in \eqref{obj_main} is to maximize the minimum uplink throughput among all UDs. Accordingly, we define the instantaneous reward $r_t \in \mathcal{R}$ as
\begin{equation}
r_t = \min_{k} R^{ul}_{k,t}.
\label{eq:reward_utility}
\end{equation}

To incorporate the downlink-throughput and UAV-energy requirements in~\eqref{c0} and~\eqref{c7} into the primal-dual learning framework, we define two nonnegative instantaneous constraint costs as
\begin{align}
c^{dl}_t &= \frac{\big[R_{\min}-\min_{k\in\mathcal{K}}R^{dl}_{k,t}\big]_+}{R_{\min}}, \label{eq:cost_dl}\\
c^{uav}_t &= \frac{\big[E_{th}-(E_{bat}-E_{uav,t})\big]_+}{E_{bat}}. \label{eq:cost_c7}
\end{align}
Herein, $c^{dl}_t$ quantifies the normalized violation of the downlink throughput requirement, whereas $c^{uav}_t$ quantifies the normalized violation of the UAV residual-energy constraint. By construction, each cost is $0$ if and only if its corresponding per-frame constraint is satisfied. These cost signals are stored together with $r_t$ in the replay buffer and are used to train the corresponding cost critics and update the associated dual variables, as described in Sections~\ref{PDUpdateSec} and~\ref{DualUpdateSec}, thereby guiding the learned policy toward satisfying the original per-frame requirements. Since $r_t$ constitutes a lower bound on the uplink throughput of all UDs in each frame, maximizing $r_t$ ensures max-min fairness among UDs, which is essential in IoUT applications requiring comprehensive spatial monitoring, such as smart agriculture. When other performance metrics are prioritized, e.g., reliability and latency in underground pipeline leakage detection or energy efficiency in long-term border patrol~\cite{WUSNReview, IoUTReview, LinSAGUIN}, the proposed constrained MDP formulation readily accommodates them by tailoring the reward and constraint costs accordingly.
\subsection{State Transition}
\label{subsec:transition}
At decision epoch $t$, the UAV selects the resource-control action $\mathbf{a}_t$ based on the current state $\mathbf{s}_t$. The resulting resource allocation determines the communication performance and UAV energy expenditure for the corresponding frame, yielding $r_t$ and $\bar{E}_{t+1}$. After the frame evaluation, the environment advances to the next daily VWC sample and generates new small-scale fading realizations for the downlink and uplink channels, thereby forming the next state $\mathbf{s}_{t+1}$. Accordingly, the state transition is represented by $\mathbf{s}_{t+1}\sim\mathcal{P}(\cdot|\mathbf{s}_t,\mathbf{a}_t)$. Hence, the previous action influences the subsequent state through the feedback components $r_t$ and $\bar{E}_{t+1}$, which the actor takes as inputs together with the updated channel and soil conditions when making the next decision. Since the soil-moisture dynamics and small-scale fading are difficult to model analytically, $\mathcal{P}$ is unknown and site-specific, and the MDP formulation enables the agent to learn an adaptive resource-allocation policy directly from trace-driven interactions. Moreover, this formulation can directly incorporate stronger inter-frame coupling, such as cumulative UAV battery depletion and UAV trajectory design, by augmenting the state and constraint costs.

\section{Proposed PD-TQC-based DRL Solution}
\label{sec:framework}
\begin{figure}[!t]
\centering
\includegraphics[width=3.45in]{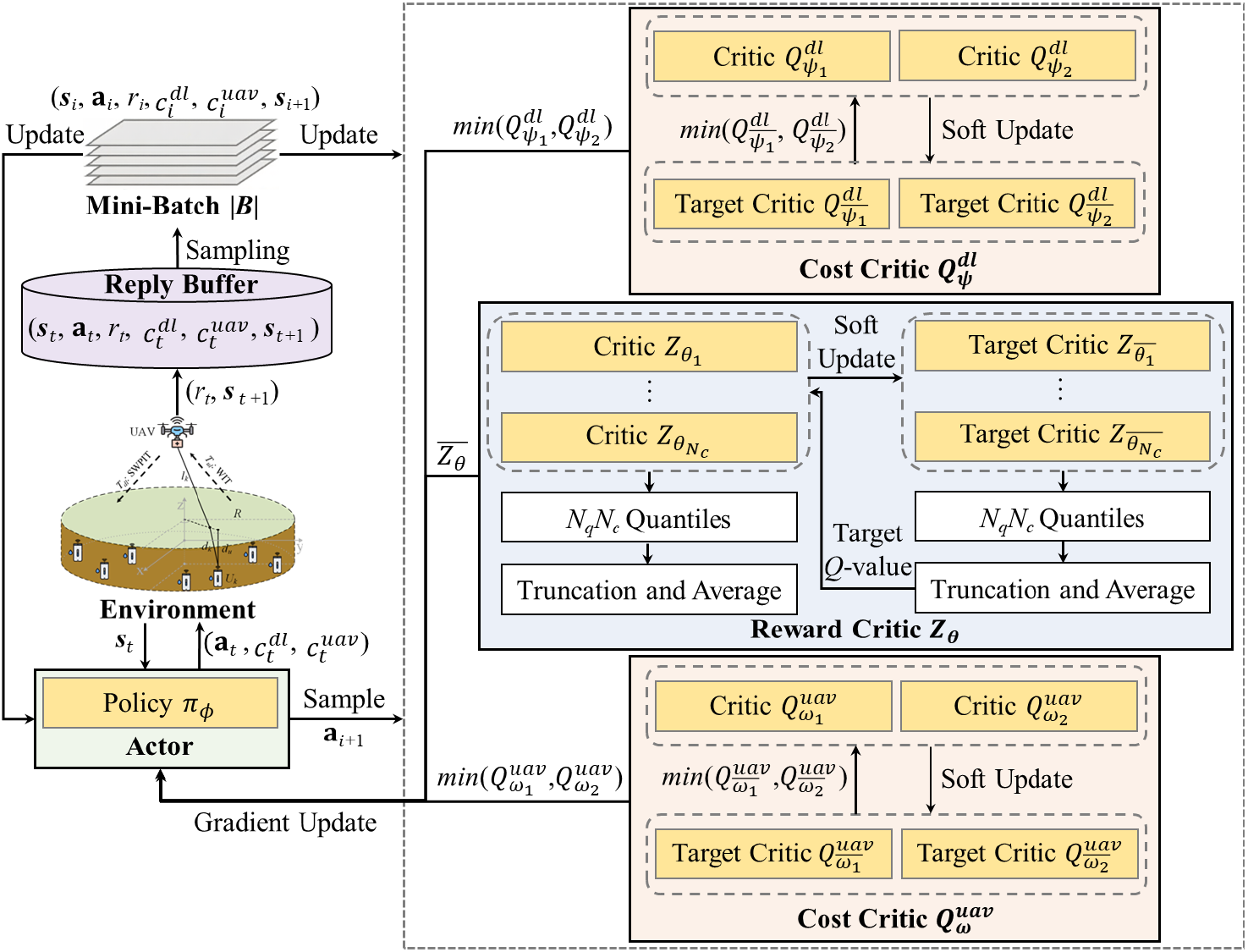}
\caption{An illustration of the PD-TQC algorithm framework for the considered UAV-enabled WPUSN system.}
\label{fig_drlarch}
\end{figure}

To address the coupled and non-convex resource allocation problem in \eqref{obj_main}--\eqref{c8} under stochastic fading and time-varying soil conditions, inspired by~\cite{PDPPO, PDSAC, TQCAlgorithm}, we propose an off-policy PD-TQC framework that adopts an entropy-regularized actor-critic paradigm with experience replay, as illustrated in Fig.~\ref{fig_drlarch}. At each daily decision epoch $t$, the UAV observes the state $\mathbf{s}_t$ in \eqref{eq:state_def_new} and samples a continuous action $\mathbf{a}_t \sim \pi_{\phi}(\cdot|\mathbf{s}_t)$ from a stochastic actor parameterized by $\phi$. After executing $\mathbf{a}_t$, the environment returns the reward $r_t$ in \eqref{eq:reward_utility}, two constraint-cost signals $(c^{dl}_t,c^{uav}_t)$ in \eqref{eq:cost_dl} and \eqref{eq:cost_c7}, and the next state $\mathbf{s}_{t+1}$. The transition tuple $(\mathbf{s}_t,\mathbf{a}_t,r_t,c^{dl}_t,c^{uav}_t,\mathbf{s}_{t+1})$ is stored in a replay buffer $\mathcal{D}$.

During training, mini-batches $\mathcal{B}$ are randomly sampled from $\mathcal{D}$, i.e., $\{(\mathbf{s}_i,\mathbf{a}_i,r_i,c^{dl}_i,c^{uav}_i,\mathbf{s}_{i+1})\}_{i=1}^{|\mathcal{B}|}$, where $i$ indexes sampled replay transitions rather than temporally adjacent decision epochs. For each mini-batch, PD-TQC performs three coupled updates aligned with Fig.~\ref{fig_drlarch}: (i) it updates a set of distributional reward critics using TQC to obtain truncated distributional return estimates; (ii) it updates two pairs of twin cost critics via clipped double Q-learning to estimate discounted constraint-violation returns; and (iii) it updates the actor by minimizing an entropy-regularized Lagrangian objective constructed from the learned reward and cost critics. The dual variables are then updated according to the average constraint costs observed during training, and all target critics are softly updated via Polyak averaging. Notably, for each sampled transition, the same stop-gradient next action $\mathbf{a}_{i+1}\sim\pi_{\phi}(\cdot|\mathbf{s}_{i+1})$ is reused to form the bootstrap targets of both reward and cost critics, ensuring consistent off-policy bootstrapping across modules.

\subsection{TQC-Based Reward Critic}
To address the pronounced reward fluctuations induced by small-scale fading and time-varying soil conditions in the considered UAV-enabled WPUSN, we adopt the TQC algorithm to learn the reward critic, which extends the standard SAC framework with distributional value estimation and quantile truncation~\cite{TQCAlgorithm}. Unlike SAC, which relies on scalar Q-value estimates, TQC leverages distributional DRL to explicitly characterize return uncertainty. By approximating the return distribution through quantile critics rather than relying solely on expected action values, TQC is particularly effective in environments with highly stochastic returns and time-varying channel dynamics, rendering it well suited to the underground monitoring scenario.

The key innovation of TQC lies in its distributional value representation and truncated Bellman backup mechanism. Instead of maintaining point estimates of Q-values as in SAC, TQC employs an ensemble of quantile critics to approximate the return distribution via quantile regression. This distributional modeling captures the inherent uncertainty arising from underground channel variability and provides a richer and more informative training signal for policy learning. Furthermore, by discarding the highest quantiles during target construction, TQC systematically suppresses Q-value overestimation, resulting in more stable and reliable policy updates.

Mathematically, TQC models the action-value distribution via quantile regression. For a given state-action pair $(\mathbf{s}_i,\mathbf{a}_i)$, each critic outputs $N_q$ quantile estimates.
Specifically, we maintain an ensemble of $N_c$ reward critics parameterized by $\{\theta_j\}_{j=1}^{N_c}$, i.e.,
\begin{equation}
Z_{\theta_j}(\mathbf{s}_i,\mathbf{a}_i) = \big[z_{\theta_j}^{(1)}(\mathbf{s}_i,\mathbf{a}_i),\ldots, z_{\theta_j}^{(N_q)}(\mathbf{s}_i,\mathbf{a}_i) \big]\in\mathbb{R}^{N_q},
\label{eq:tqc_quantile_out}
\end{equation}
where $z_{\theta_j}^{(m)}(\mathbf{s}_i,\mathbf{a}_i)$ approximates the $\hat{\tau}_m$-quantile with fixed fractions
$\hat{\tau}_m=(m-0.5)/N_q$, $m=1,\ldots,N_q$.

For each sampled transition $(\mathbf{s}_i,\mathbf{a}_i,r_i,\mathbf{s}_{i+1})$, we sample
$\mathbf{a}_{i+1}\sim\pi_{\phi}(\cdot|\mathbf{s}_{i+1})$ and compute $\log\pi_{\phi}(\mathbf{a}_{i+1}|\mathbf{s}_{i+1})$.
Then, we aggregate the target quantiles predicted by the target ensemble:
\begin{equation}
\mathcal{Z}'_{i+1} = \bigcup_{j=1}^{N_c} Z_{\bar{\theta}_j}(\mathbf{s}_{i+1},\mathbf{a}_{i+1}) \in\mathbb{R}^{N_cN_q},
\end{equation}
where $\{\bar{\theta}_j\}$ are the Polyak-averaged target parameters.
Following TQC, we sort $\mathcal{Z}'_{i+1}$ and discard the largest $N_{tr}$ elements, i.e.,
\begin{equation}
\tilde{\mathcal{Z}}'_{i+1}=\mathrm{Trunc}_{N_{tr}}\!\left(\mathrm{Sort}\!\left(\mathcal{Z}'_{i+1}\right)\right)\in\mathbb{R}^{N_cN_q-N_{tr}},
\label{eq:tqc_trunc}
\end{equation}
where $N_{tr}\in\{0,\ldots,N_cN_q-1\}$ controls conservativeness. For instance, larger $N_{tr}$ removes high quantiles prone to optimistic bias.

Accordingly, the entropy-regularized bootstrap target vector is given by
\begin{equation}
\mathbf{y}^{(r)}_i = r_i+\gamma\!\left(\tilde{\mathcal{Z}}'_{i+1} -\alpha_{ent}\log\pi_{\phi}(\mathbf{a}_{i+1}|\mathbf{s}_{i+1}) \right),
\label{eq:tqc_target}
\end{equation}
where $\alpha_{ent}>0$ is the entropy temperature. Since $r_i$ corresponds to the per-frame max-min objective in~\eqref{obj_main}, the reward critics estimate discounted returns of per-frame minimum uplink throughputs. As the channel realizations are exogenous and no energy state is carried across frames, $\mathbf{a}_i$ affects only the current-frame reward and costs, rendering the expected future-value term in~\eqref{eq:tqc_target} independent of $\mathbf{a}_i$. Hence, for any $\gamma\in[0, 1)$, maximizing the discounted return also maximizes the expected per-frame objective, while the discounted formulation remains applicable to extensions with inter-frame coupling, such as cumulative UAV battery depletion or UAV trajectory design.

Each reward critic is updated by minimizing the quantile Huber loss, which is given by
\begin{equation}
\mathcal{L}_{r}(\theta_j) = \mathbb{E}\!\left[ \frac{1}{N_q\,|\mathbf{y}^{(r)}_i|} \sum_{m=1}^{N_q} \sum_{n=1}^{|\mathbf{y}^{(r)}_i|} \rho_{\hat{\tau}_m}^{\kappa_h}
\Big( y^{(r)}_{i,n}-z_{\theta_j}^{(m)}(\mathbf{s}_i,\mathbf{a}_i) \Big)
\right],
\label{eq:quantile_huber_loss}
\end{equation}
where $|\mathbf{y}^{(r)}_i|=N_cN_q-N_{tr}$, $\kappa_h$ is the Huber threshold, while $\rho_{\hat{\tau}}^{\kappa_h}(\cdot)$ is defined in~\cite{TQCAlgorithm}.

During the actor update, the reward action-value estimate used for policy improvement is obtained by sorting, truncating, and averaging the predicted quantiles, i.e.,
\begin{equation} 
\bar{Z}_{\theta}(\mathbf{s}_i,\mathbf{a}_i) = \mathrm{Mean} \left( \mathrm{Trunc}_{N_{tr}} \Big( \mathrm{Sort} \big( \bigcup_{j=1}^{N_c} Z_{\theta_j}(\mathbf{s}_i,\mathbf{a}_i) \big) \Big) \right). \label{eq:tqc_util_est} 
\end{equation} 
This truncated statistic serves as the primal reward term in the Lagrangian actor objective. The same sampled $\mathbf{a}_{i+1}$ is reused to form temporal difference (TD) targets for the cost critics, ensuring consistent bootstrapping across reward and constraint learning.

\subsection{Twin Cost Critics and Actor} \label{PDUpdateSec}
To incorporate the per-frame constraints into policy learning, we learn cost critics that approximate the expected discounted constraint costs. Specifically, we use the instantaneous costs in~\eqref{eq:cost_dl} and~\eqref{eq:cost_c7} to represent the downlink-throughput requirement~\eqref{c0} and the UAV residual-energy requirement~\eqref{c7}, respectively. These cost signals are stored together with the reward in the replay buffer to learn cost-to-go functions.

For each cost, we employ twin cost critics and use the minimum of their target estimates for TD bootstrapping, yielding $\{Q^{dl}_{\psi_1},Q^{dl}_{\psi_2}\}$ and $\{Q^{uav}_{\omega_1},Q^{uav}_{\omega_2}\}$. For each sampled transition $(\mathbf{s}_i,\mathbf{a}_i,r_i,c^{dl}_i,c^{uav}_i,\mathbf{s}_{i+1})$, we reuse the same next action $\mathbf{a}_{i+1}\sim\pi_{\phi}(\cdot|\mathbf{s}_{i+1})$ as in \eqref{eq:tqc_target} to construct consistent TD targets, i.e.,
\begin{align} 
y^{dl}_i &= c^{dl}_i+\gamma\min_{j\in\{1,2\}}Q^{dl}_{\bar{\psi}_j}(\mathbf{s}_{i+1},\mathbf{a}_{i+1}), \label{eq:td_dl} \\
y^{uav}_i &= c^{uav}_i+\gamma\min_{j\in\{1,2\}}Q^{uav}_{\bar{\omega}_j}(\mathbf{s}_{i+1},\mathbf{a}_{i+1}). \label{eq:td_uav} 
\end{align}
Each cost critic is updated by minimizing the mean-squared TD error, which is given by
\begin{align}
\mathcal{L}^{dl}_{Q}(\psi_j) 
&= \mathbb{E}\!\left[\big(Q^{dl}_{\psi_j}(\mathbf{s}_i,\mathbf{a}_i)-y^{dl}_i\big)^2\right], \quad j\in\{1,2\}, \label{eq:pd_tqc_cost_mse_dl} \\
\mathcal{L}^{uav}_{Q}(\omega_j) 
&= \mathbb{E}\!\left[\big(Q^{uav}_{\omega_j}(\mathbf{s}_i,\mathbf{a}_i)-y^{uav}_i\big)^2\right], \quad j\in\{1,2\}. \label{eq:pd_tqc_cost_mse_uav}
\end{align}
With the reward and cost critics updated, the actor samples a new action $\mathbf{a}^{\pi}_i\sim\pi_{\phi}(\cdot|\mathbf{s}_i)$ for each sampled state and evaluates it using the learned reward and cost critics. The truncated reward estimate is given by $\bar{Z}_{\theta}(\mathbf{s}_i,\mathbf{a}^{\pi}_i)$ in~\eqref{eq:tqc_util_est}, while the corresponding cost estimates are defined as $Q^{dl}_{\min}(\mathbf{s}_i,\mathbf{a}^{\pi}_i)=\min\!\big(Q^{dl}_{\psi_1}(\mathbf{s}_i,\mathbf{a}^{\pi}_i),Q^{dl}_{\psi_2}(\mathbf{s}_i,\mathbf{a}^{\pi}_i)\big)$ and $Q^{uav}_{\min}(\mathbf{s}_i,\mathbf{a}^{\pi}_i)=\min\!\big(Q^{uav}_{\omega_1}(\mathbf{s}_i,\mathbf{a}^{\pi}_i),Q^{uav}_{\omega_2}(\mathbf{s}_i,\mathbf{a}^{\pi}_i)\big)$. Based on these estimates, the actor parameters are updated by minimizing the following entropy-regularized Lagrangian objective, i.e.,
\begin{align}
\mathcal{L}_{\pi}(\phi) = \mathbb{E}\Big[
\alpha_{ent}\log\pi_{\phi}(\mathbf{a}^{\pi}_i|\mathbf{s}_i)
&- \bar{Z}_{\theta}(\mathbf{s}_i,\mathbf{a}^{\pi}_i)
+ \lambda_{dl}Q^{dl}_{\min}(\mathbf{s}_i,\mathbf{a}^{\pi}_i) \nonumber\\
&+\lambda_{uav}Q^{uav}_{\min}(\mathbf{s}_i,\mathbf{a}^{\pi}_i)
\Big],
\label{eq:actor_loss_tqc_lag}
\end{align}
where $\lambda_{dl} \ge 0$ and $\lambda_{uav} \ge 0$ are the Lagrange multipliers associated with the constraint costs corresponding to~\eqref{c0} and~\eqref{c7}, respectively, balancing reward maximization against the corresponding constraint-cost returns.

\subsection{Dual Variables and Target Network Updates} \label{DualUpdateSec}
Finally, we update the dual variables to account for the downlink throughput and UAV residual-energy requirements during policy learning. In our implementation, the dual variables are updated once per training epoch based on the epoch-average costs $\overline{c}^{dl}$ and $\overline{c}^{uav}$, i.e.,
\begin{align}
\lambda_{dl} &\leftarrow \Pi_{[0,\lambda_{\max}]}\Big(\lambda_{dl}+\eta_{\lambda}(\overline{c}^{dl}-d_{dl})\Big), \label{Equpdatedl}\\
\lambda_{uav} &\leftarrow \Pi_{[0,\lambda_{\max}]}\Big(\lambda_{uav}+\eta_{\lambda}(\overline{c}^{uav}-d_{uav})\Big),
\label{Equpdateuav}
\end{align}
where $d_{dl}$ and $d_{uav}$ denote the reference levels for the epoch-average constraint costs and are set to $0$, $\eta_{\lambda}>0$ is the dual step size, and $\Pi_{[0,\lambda_{\max}]}(\cdot)$ is the projection onto $[0,\lambda_{\max}]$. Since both constraint costs are nonnegative, zero epoch-average costs imply satisfaction of the corresponding per-frame constraints. Setting their reference levels to $0$ therefore preserves the original feasibility requirements as optimization targets, while the dual updates increase the penalties for constraint violations to guide the learned policy toward these targets. Importantly, since $\overline{c}^{dl}$ and $\overline{c}^{uav}$ are computed from the observed costs rather than the cost-critic estimates, the dual updates are not affected by the potential cost underestimation induced by the minimum operator in~\eqref{eq:td_dl}, \eqref{eq:td_uav}, and~\eqref{eq:actor_loss_tqc_lag}~\cite{CALSafeRL}. Instead, any constraint violation resulting from such underestimation increases the corresponding multiplier and is thus penalized more strongly in subsequent policy updates.

To stabilize training, the target networks for the critics,
i.e., $\{Z_{\bar{\theta}_j}, Q^{dl}_{\bar{\psi}_j}, Q^{uav}_{\bar{\omega}_j}\}$,
are softly updated via Polyak averaging. Given an arbitrary online/target parameter pair $(\vartheta,\bar{\vartheta})$, the soft update is given by
\begin{equation}
\bar{\vartheta}\leftarrow (1-\zeta)\bar{\vartheta}+\zeta\,\vartheta,
\label{eq:pd_tqc_polyak_update}
\end{equation}
where $\zeta\in(0,1)$ is the update factor.

Algorithm~\ref{alg:pd_tqc_lag} outlines the proposed PD-TQC framework. By integrating the TQC-based reward evaluation with twin cost critics, PD-TQC optimizes the max-min throughput objective while accounting for the per-frame downlink throughput and UAV residual-energy requirements. The stochastic actor outputs continuous resource-allocation decisions, and the dual variables are adaptively updated to enforce constraint satisfaction rather than relying on fixed penalty coefficients. Moreover, the framework can scale to multiple constraints by incorporating additional cost critics and the corresponding Lagrange multipliers.

\begin{algorithm}[t]
\caption{PD-TQC Algorithm for UAV-Enabled WPUSN Systems}
\label{alg:pd_tqc_lag}
\normalsize
\begin{algorithmic}[1]
\REQUIRE Discount factor $\gamma$; truncation parameter $N_{tr}$; entropy temperature $\alpha_{ent}$; ensemble size $N_c$ and quantile count $N_q$; Huber threshold $\kappa_h$; learning rates $(\eta_{\pi},\eta_Z,\eta_Q)$; dual step size $\eta_{\lambda}$; target update factor $\zeta$; dual-variable upper bound $\lambda_{\max}$; constraint-cost reference levels $(d_{dl},d_{uav})$; mini-batch size $|\mathcal{B}|$; epoch length $N_{day}$ and number of epochs $N_{ep}$.
\STATE Initialize stochastic actor $\pi_{\phi}$.
\STATE Initialize reward critics $\{Z_{\theta_j}\}_{j=1}^{N_c}$ and targets $\{Z_{\bar{\theta}_j}\leftarrow Z_{\theta_j}\}_{j=1}^{N_c}$.
\STATE Initialize twin cost critics $\{Q^{dl}_{\psi_1},Q^{dl}_{\psi_2}\}$, $\{Q^{uav}_{\omega_1},Q^{uav}_{\omega_2}\}$ and their targets $\{Q^{dl}_{\bar{\psi}_j}\leftarrow Q^{dl}_{\psi_j}\}$, $\{Q^{uav}_{\bar{\omega}_j}\leftarrow Q^{uav}_{\omega_j}\}$ for $j\in\{1,2\}$.
\STATE Initialize replay buffer $\mathcal{D}$ and set $\lambda_{dl},\lambda_{uav} \ge 0$.
\FOR{epoch $e=1$ to $N_{ep}$}
    \STATE Reset environment and obtain initial state $\mathbf{s}_t$.
    \STATE Initialize epoch cost lists $\mathcal{C}_{dl}\leftarrow\emptyset$, $\mathcal{C}_{uav}\leftarrow\emptyset$.
    \FOR{decision epoch $t=1$ to $N_{day}$}
        \STATE Sample $\mathbf{a}_t\sim\pi_{\phi}(\cdot|\mathbf{s}_t)$.
        \STATE Execute $\mathbf{a}_t$, compute the uplink transmit powers via \eqref{eq:ul_clip}, and observe $(r_t,c^{dl}_t,c^{uav}_t,\mathbf{s}_{t+1})$.
        \STATE Store $(\mathbf{s}_t,\mathbf{a}_t,r_t,c^{dl}_t,c^{uav}_t,\mathbf{s}_{t+1})$ into $\mathcal{D}$.
        \STATE Append $c^{dl}_t$ to $\mathcal{C}_{dl}$ and $c^{uav}_t$ to $\mathcal{C}_{uav}$.
        \IF{$|\mathcal{D}|\ge|\mathcal{B}|$}
            \STATE Sample a mini-batch $\{(\mathbf{s}_i,\mathbf{a}_i,r_i,c^{dl}_i,c^{uav}_i,\mathbf{s}_{i+1})\}_{i=1}^{|\mathcal{B}|}$ from $\mathcal{D}$.
            \STATE Sample $\mathbf{a}_{i+1}\sim\pi_{\phi}(\cdot|\mathbf{s}_{i+1})$ with stop-gradient.
            \STATE Compute $\tilde{\mathcal{Z}}'_{i+1}$ and $\mathbf{y}^{(r)}_i$ via \eqref{eq:tqc_trunc} and \eqref{eq:tqc_target}.
            \STATE Update $\{Z_{\theta_j}\}$ by minimizing \eqref{eq:quantile_huber_loss}.
            \STATE Construct $y^{dl}_i$ and $y^{uav}_i$ via \eqref{eq:td_dl} and \eqref{eq:td_uav}, and update the twin cost critics by minimizing \eqref{eq:pd_tqc_cost_mse_dl} and \eqref{eq:pd_tqc_cost_mse_uav}.
            \STATE Sample $\mathbf{a}^{\pi}_i\sim\pi_{\phi}(\cdot|\mathbf{s}_i)$, evaluate $\bar{Z}_{\theta}(\mathbf{s}_i,\mathbf{a}^{\pi}_i)$ and the corresponding cost estimates, and update the actor by minimizing \eqref{eq:actor_loss_tqc_lag}.
            \STATE Soft-update target critics $\{Z_{\bar{\theta}_j}, Q^{dl}_{\bar{\psi}_j}, Q^{uav}_{\bar{\omega}_j}\}$ using \eqref{eq:pd_tqc_polyak_update} with factor $\zeta$.
        \ENDIF
        \STATE $\mathbf{s}_t\leftarrow\mathbf{s}_{t+1}$.
    \ENDFOR
    \STATE Compute epoch-average costs $\overline{c}^{dl}=\mathrm{mean}(\mathcal{C}_{dl})$ and $\overline{c}^{uav}=\mathrm{mean}(\mathcal{C}_{uav})$.
    \STATE Update $(\lambda_{dl},\lambda_{uav})$ using \eqref{Equpdatedl} and \eqref{Equpdateuav}.
\ENDFOR
\ENSURE Use the learned $\pi_{\phi}$ for online control.
\end{algorithmic}
\end{algorithm}

\subsection{Computational Complexity}
\label{subsec:complexity}
We analyze the computational complexity of PD-TQC by distinguishing the offline training phase from the online inference phase. Let $N_1$ and $N_2$ denote the numbers of neurons in the two hidden layers, and let $|\phi|$, $|\theta_j|$, and $|\psi_j|=|\omega_j|$ denote the numbers of parameters of the actor, each reward critic, and each cost critic, respectively. In the offline training phase, each epoch consists of $N_{day}$ environment interactions and $N_{day}$ gradient updates, where each update samples a mini-batch $\mathcal{B}$ from the replay buffer $\mathcal{D}$ to train the actor, $N_c$ reward critics, and four cost critics. Hence, the training complexity is $\mathcal{O}\big(N_{ep}N_{day}|\mathcal{B}|(|\phi|+N_c|\theta_j|+4|\psi_j|+N_cN_q\log(N_cN_q))\big)$. In the online inference phase, only the actor is deployed at the UAV, and each decision requires a single forward pass with complexity $\mathcal{O}\big((4K+3)N_1+N_1N_2+5(K+1)N_2\big)$. Therefore, the computationally intensive learning is performed offline before deployment, while the online decision-making avoids repeatedly solving the non-convex problem in~\eqref{obj_main}--\eqref{c8}. The measured training time and inference latency of all schemes are compared in Section~\ref{sec:res}.

\section{Performance Evaluation}\label{sec:res}
This section presents the key simulation configurations and demonstrates the performance of the proposed PD-TQC approach for solving the formulated optimization problem.

\subsection{Simulation Configuration}
In this simulation, we consider a practical UAV-enabled WPUSN system for a real-world center-pivot irrigation farm located in Monahans, Texas, USA, as the study scenario. Unless stated otherwise, the UAV is equipped with $Q=8$ antenna elements and hovers at $H=5$~m above the ground while $K=7$ UDs are uniformly and randomly deployed within a circular area of radius $R=50$~m and buried at a depth of $d_u = 0.5$~m. This burial depth is selected as a representative shallow deployment to mitigate disturbances from conventional agricultural operations while maintaining feasible SWIPT operation under the considered system conditions. To ensure realistic modeling, the \textit{in-situ} soil clay percentage and daily VWC are obtained from~\cite{Undergroundfield, VWCdata} to accurately characterize attenuation in soil. Herein, the DRL agent is trained using the full-year 2024 daily VWC sequence, while the full-year 2025 VWC record is reserved for out-of-sample evaluation. Each decision epoch is indexed by one daily-mean VWC sample, which updates the soil-dependent attenuation for the corresponding communication frame, while small-scale fading is independently generated at each decision epoch. Following previous studies~\cite{LinBSWPUSNs, LinCSIfree, LinRIS, LinUAV}, we adopt a Rician fading factor of $\kappa_R=10$~dB and an aboveground path-loss exponent of $\varsigma=2$ to characterize reference propagation conditions with a dominant signal component. Such a dominant component is plausible for the considered air-soil links, since the UAV hovers above open farmland, while the large permittivity contrast at the air-soil interface refracts the signal nearly perpendicularly into the soil and multipath components traversing longer underground paths experience considerably stronger soil attenuation~\cite{Undergroundfield, LinBSWPUSNs}. These settings allow us to focus on the joint resource optimization of the proposed RS-SWIPT framework. Since soil heterogeneity and surface vegetation may weaken this dominant component, the impact of $\kappa_R$ is further evaluated in Section~\ref{sec:res}-B. The channel model and its parameters can be further refined based on site-specific propagation conditions to obtain more realistic estimates of system performance in practical deployments. The other key environmental parameters are listed in~Table~\ref{tab1}. 

For the proposed PD-TQC algorithm, the actor and critic networks are implemented as fully connected neural networks with two hidden layers of $512$ and $256$ neurons, respectively. The actor applies layer normalization followed by ReLU activation after each hidden layer, whereas the reward and cost critics employ ReLU activations without layer normalization. All networks are optimized using AdamW with a learning rate of $10^{-4}$. The reward critic consists of an ensemble of $N_c=5$ quantile critics, each predicting $N_q=50$ return quantiles. To reduce overestimation, we construct the target by discarding the largest $N_{tr}=10$ quantiles (i.e., TQC truncation) and train the critics using quantile Huber regression with a threshold of $\kappa_h=1$. The entropy temperature is fixed at $\alpha_{ent}=0.1$, and all target networks are updated via Polyak averaging with an update factor $\zeta=0.005$. The remaining training hyper-parameters are summarized in Table~\ref{tab1}.

\begin{table}[t]
\caption{Simulation Parameters}
\label{tab1}
\centering
\renewcommand{\arraystretch}{1}
\begin{tabular}{m{0.58\columnwidth}<{\raggedright} m{0.3\columnwidth}<{\centering}}
\toprule
\textbf{Parameters} & \textbf{Values} \\ \midrule

\multicolumn{2}{l}{\textbf{Operating Scenario}~\cite{Undergroundfield, VWCdata, LinUAV, LinBSWPUSNs}} \\ \midrule
Number of UDs ($K$) & $7$ \\
Number of UAV antennas ($Q$) & $8$ \\
Deployment radius ($R$) & $50$~m \\
UAV altitude ($H$) & $5$~m \\
Frame duration ($T$) & $1$~s \\
Bandwidth ($B$) & $125$~kHz \\
Carrier frequency ($f$) & $433$~MHz \\
Noise power ($\sigma^2$) & $-117$~dBm \\
Rician factor for downlink/uplink links ($\kappa_R$) & $10$~dB \\
Air path-loss exponent ($\varsigma$) & $2$ \\
UAV antenna gain ($G_{uav}$) & $5$~dBi \\
UD antenna gain ($G_{k}$) & $2.15$~dBi \\
Max UAV transmit power ($P_{\max}$) & $30$~dBm \\
Burial depth of UDs ($d_u$) & $0.5$~m \\
Clay percentage ($m_c$) & $38\%$ \\
Daily-mean VWC  &Trace from~\cite{VWCdata} \\
Nonlinear EH saturation power ($M$) & $24$~mW \\
EH circuit constants ($a,b$) & $a=150,\; b=0.0014$ \\
Uplink transmit-energy fraction ($\varphi_k$) & $0.6$  \\
Per-UD maximum transmit power ($P_{k,\max}$) & $14$~dBm \\
Downlink throughput threshold ($R_{\min}$) & $20$~kbps \\
UAV hovering power ($P_{hov}$) & $56.2926$~W \\
Per-frame UAV energy budget ($E_{bat}$) & $67$~J \\
Residual UAV energy threshold ($E_{th}$) & $10.05$~J \\

\midrule
\multicolumn{2}{l}{\textbf{PD-TQC Configuration}} \\ 
\midrule
Decision epochs per training epoch ($N_{day}$) & $366$ \\
Number of epochs ($N_{ep}$) & $1200$ \\
Discount factor ($\gamma$) & $0.8$ \\
Mini-batch size ($|\mathcal{B}|$) & $256$ \\
Replay buffer size & $10^6$ \\
Updates per trigger & $1$ \\
Number of quantile critics ($N_c$) & $5$ \\
Quantiles per critic ($N_q$) & $50$ \\
Truncation parameter ($N_{tr}$) & $10$  \\
Huber threshold ($\kappa_h$) & $1$  \\
First hidden-layer size ($N_1$) & $512$ \\
Second hidden-layer size ($N_2$) & $256$ \\
Optimizer & AdamW \\
Learning rate ($\eta_{\pi},\eta_Z,\eta_Q$) & $10^{-4}$ \\
Dual learning rate ($\eta_{\lambda}$) & $0.1$ \\
Entropy temperature ($\alpha_{ent}$) & $0.1$ \\
Polyak averaging factor ($\zeta$) & $0.005$ \\
Dual-variable upper bound ($\lambda_{\max}$) & $30$ \\
\bottomrule
\end{tabular}
\renewcommand{\arraystretch}{0.9}
\end{table}

For performance comparison, four benchmark schemes are implemented:
\begin{enumerate}
    \item \textbf{Greedy}~\cite{GreedyApp}: As a learning-free baseline, we implement a Greedy strategy that selects the action maximizing the minimum uplink throughput while penalizing the downlink throughput-threshold violation and the UAV residual-energy constraint violation.

    \item \textbf{PPO}~\cite{PDPPO}: An on-policy actor-critic approach that updates the policy and value function by maximizing a clipped surrogate objective. The downlink throughput and UAV residual-energy constraints are enforced via primal-dual updates of the associated Lagrange multipliers.

    \item \textbf{SAC}~\cite{PDSAC}: An off-policy actor-critic mechanism that learns a stochastic policy by maximizing the expected entropy-regularized return. The objective is learned using twin reward critics, whereas the constraints are handled using separate twin cost critics together with epoch-averaged primal-dual updates.

    \item \textbf{PD-TQC with SDMA-based SWIPT}: The proposed PD-TQC algorithm is employed to control the SDMA-based SWIPT downlink~\cite{SDMAApp} and the TDMA-based uplink in the considered UAV-enabled WPUSN system.
\end{enumerate}
Note that both the PPO and SAC baselines adopt the same primal-dual Lagrange multiplier update mechanism as the proposed approach to ensure a fair comparison under an identical constraint-handling framework.

\subsection{Numerical Results Analysis}
\subsubsection{Algorithm Performance Comparison}
\begin{figure}[!t]
\centering
\includegraphics[width=3.4in]{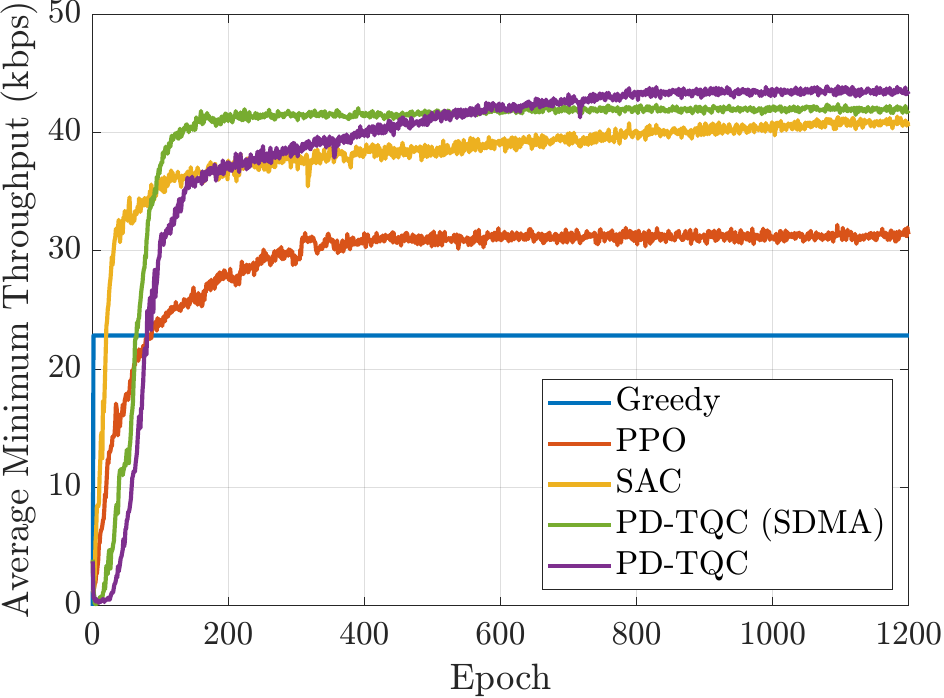}
\caption{Training performance of the Greedy, PPO, SAC, and PD-TQC approaches in the UAV-enabled WPUSN system using the 2024 annual VWC data, as well as the PD-TQC scheme with SDMA-based SWIPT.}
\label{fig_conres}
\end{figure}

We first present the reward results (i.e., the annual average of the minimum uplink throughput across all UDs) during the training process for the Greedy, PPO, SAC, and PD-TQC approaches in the UAV-enabled WPUSN system using the 2024 annual VWC dataset, as well as a PD-TQC baseline using SDMA-based SWIPT, as shown in Fig.~\ref{fig_conres}. One can observe that the Greedy scheme does not undergo iterative policy training and therefore appears as a fixed performance reference, whereas the other approaches require more epochs to gradually balance exploration and exploitation and achieve stable policy performance. Specifically, PPO, SAC, and PD-TQC with SDMA-based SWIPT converge after approximately $600$, $1000$, and $500$ epochs, respectively, whereas the proposed PD-TQC approach attains the highest steady-state annual average minimum uplink throughput of $43.47$~kbps by $800$ epochs. Since SDMA is a special case of RSMA with $\beta_c=0$ in~\eqref{EqTxSignal}~\cite{SDMAApp, MaoRSMAsurvey}, every feasible SDMA-based allocation is also feasible under RSMA. Moreover, the two PD-TQC schemes adopt the same learning algorithm, training and test datasets, PS receivers, and TDMA-based uplink. Therefore, although the SDMA-based scheme converges faster owing to its smaller state and action spaces, its lower saturation level indicates that the performance gap stems from the additional flexibility of RSMA exploited by the learned policy rather than from insufficient training. Upon convergence, the proposed PD-TQC achieves an annual average minimum uplink throughput that is $90.28\%$, $38.89\%$, $6.45\%$, and $3.66\%$ higher than those of the Greedy, PPO, SAC, and PD-TQC with SDMA-based SWIPT schemes, respectively.

Table~\ref{tab1violate} summarizes the evaluation results of the Greedy, PPO, SAC, PD-TQC with SDMA-based SWIPT, and proposed PD-TQC schemes in the considered UAV-enabled WPUSN system under the 2025 annual VWC dataset, including the annual average minimum uplink and downlink throughputs across all UDs, as well as the per-frame satisfaction rates of the downlink throughput and UAV residual-energy constraints, i.e., the percentages of test frames in which the corresponding constraints are satisfied. The reported values are averaged over $100$ test runs with independent small-scale fading seeds for the learning-based schemes, while the Greedy benchmark is evaluated over a single $365$-frame sequence. As shown in Table~\ref{tab1violate}, the proposed PD-TQC achieves the highest minimum uplink throughput of $45.05$~kbps, yielding uplink fairness gains of $108.81\%$, $17.28\%$, $5.11\%$, and $3.99\%$ over the Greedy benchmark, PPO, SAC, and PD-TQC with SDMA-based SWIPT, respectively. Moreover, all learning-based approaches achieve annual average minimum downlink throughputs above the threshold of $R_{\min}=20$~kbps, whereas the Greedy benchmark falls below this threshold. However, meeting the threshold on average does not ensure that every UD satisfies its requirement in each frame. Note that a frame satisfies~\eqref{c0} only if the downlink throughputs of all UDs are no lower than $R_{\min}$. Although SAC achieves the highest annual average minimum downlink throughput, it satisfies the downlink throughput constraint in only $92.77\%$ of the frames, while PD-TQC with SDMA-based SWIPT satisfies it in only $85.91\%$ of the frames. In contrast, the proposed PD-TQC satisfies the downlink throughput and UAV residual-energy constraints in $96.99\%$ and $99.72\%$ of the $36{,}500$ evaluated frames, respectively, while attaining the highest minimum uplink throughput. Meanwhile, PPO achieves a higher downlink throughput constraint satisfaction rate of $99.81\%$, but its more conservative policy results in a $14.74\%$ lower minimum uplink throughput than PD-TQC.

\begin{table}[!t]
\centering
\caption{Annual average minimum uplink and downlink throughputs and per-frame constraint satisfaction rates of the considered UAV-enabled WPUSN system under the 2025 annual VWC dataset, evaluated using the trained policies of the Greedy, PPO, SAC, PD-TQC with SDMA-based SWIPT, and proposed PD-TQC schemes.}
\label{tab1violate}
\setlength{\tabcolsep}{2pt}
\renewcommand{\arraystretch}{1.1}
\begin{tabular}{m{0.24\columnwidth}<{\raggedright} m{0.16\columnwidth}<{\centering} m{0.16\columnwidth}<{\centering} m{0.16\columnwidth}<{\centering} m{0.16\columnwidth}<{\centering}}
\toprule
\textbf{Approach} & \textbf{Minimum uplink throughput (kbps)} & \textbf{Minimum downlink throughput (kbps)} & \textbf{Downlink throughput constraint satisfaction} & \textbf{UAV residual-energy constraint satisfaction} \\
\midrule
Greedy        & 21.58 & 19.81 & 53.97\% & 93.42\% \\
PPO           & 38.41 & 22.86 & \textbf{99.81\%} & 99.71\% \\
SAC           & 42.86 & \textbf{26.06} & 92.77\% & 99.61\% \\
PD-TQC (SDMA) & 43.32 & 21.08 & 85.91\% & 99.45\% \\
PD-TQC        & \textbf{45.05} & 23.54 & 96.99\% & \textbf{99.72\%} \\
\bottomrule
\end{tabular}
\end{table}

Table~\ref{tab:complexity} further compares the computational complexity of all schemes in the offline training and online inference phases. All DRL agents are trained on an NVIDIA GeForce RTX 4080 Laptop graphics processing unit, while the online decision latency is measured on an Intel Core i9-14900HX central processing unit. Although PD-TQC involves the most trainable parameters owing to its ensemble of $N_c=5$ quantile reward critics, its training converges within approximately $2.49$~h. Since this training is performed only once before deployment, it does not affect the online operation of the UAV. In the online inference phase, all DRL schemes deploy only an actor network of similar size, which requires approximately $0.31$ million floating-point operations and less than $1$~ms per decision. In contrast, the Greedy scheme evaluates $4{,}096$ candidate actions in each frame, resulting in a decision latency of $1176.66$~ms. Therefore, PD-TQC achieves the highest minimum uplink throughput among all schemes with an online decision latency of only $0.88$~ms, i.e., less than $0.1\%$ of the frame duration $T=1$~s, enabling real-time resource allocation at the UAV. Although the optimal solution of problem~\eqref{obj_main}--\eqref{c8}, approximated in each frame with perfect CSI via differential evolution~\cite{StornDE}, achieves an annual average minimum uplink throughput of $51.64$~kbps, it requires $1813.38$~ms per decision, which exceeds the frame duration and confirms the computational burden of repeated optimization.
\begin{table*}[!t]
\centering
\caption{Computational complexity of the Greedy, PPO, SAC, PD-TQC with SDMA-based SWIPT, and proposed PD-TQC schemes in the offline training and online inference phases.}
\label{tab:complexity}
\setlength{\tabcolsep}{2pt}
\renewcommand{\arraystretch}{1.1}
\begin{tabular}{m{0.15\textwidth}<{\raggedright} m{0.13\textwidth}<{\centering} m{0.13\textwidth}<{\centering} m{0.13\textwidth}<{\centering} m{0.13\textwidth}<{\centering} m{0.13\textwidth}<{\centering} m{0.13\textwidth}<{\centering}}
\toprule
\multirow[c]{2}{*}[-2.5ex]{\textbf{Approach}} & \multicolumn{3}{c}{\textbf{Offline training phase}} & \multicolumn{3}{c}{\textbf{Online inference phase}} \\
\cmidrule(lr){2-4} \cmidrule(lr){5-7}
 & \textbf{Trainable parameters} & \textbf{Training time per epoch (s)} & \textbf{Training time to convergence (h)} & \textbf{Deployed parameters} & \textbf{Floating-point operations per decision ($\times 10^{6}$)} & \textbf{Latency per decision (ms)} \\
\midrule
Greedy        & --        & --    & --   & --      & --    & 1176.66 \\
PPO           & 156,201   & 1.94  & 0.32 & 155,944 & 0.307 & 0.52 \\
SAC           & 1,117,486 & 9.98  & 2.77 & 155,944 & 0.307 & 0.78 \\
PD-TQC (SDMA) & 1,622,302 & 10.69 & 1.48 & 153,888 & 0.303 & 0.86 \\
PD-TQC        & 1,697,062 & 11.20 & 2.49 & 159,528 & 0.314 & 0.88 \\
\bottomrule
\multicolumn{7}{l}{\textit{Note: ``--'' indicates not applicable.}}
\end{tabular}
\end{table*}

\subsubsection{Impact of Rician Factor}
\begin{figure}[!t]
\centering
\includegraphics[width=3.4in]{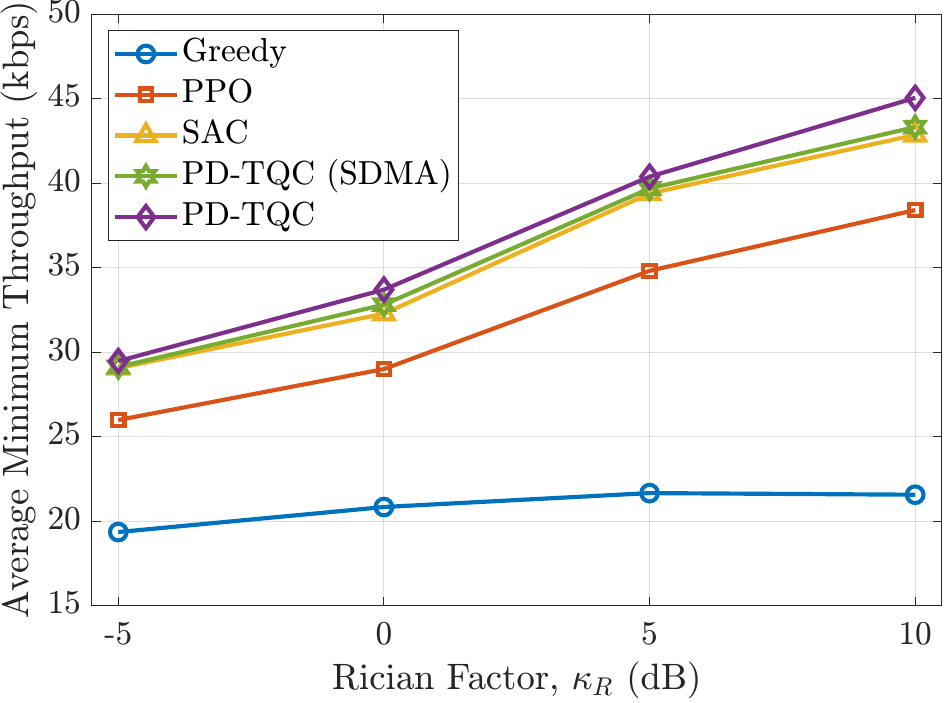}
\caption{Annual average minimum uplink throughput versus the Rician factor for the UAV-enabled WPUSN system using the 2025 annual VWC dataset, evaluated with the trained policies of the Greedy, PPO, SAC, PD-TQC with SDMA-based SWIPT, and proposed PD-TQC schemes.}
\label{fig_kres}
\end{figure}

Fig.~\ref{fig_kres} illustrates the annual average minimum uplink throughput versus the Rician factor $\kappa_R$, where a smaller $\kappa_R$ corresponds to a weaker dominant component and $\kappa_R=-5$~dB represents a scattering-dominated environment. As $\kappa_R$ decreases, the minimum uplink throughput of all learning-based schemes decreases, since more severe small-scale fading increases the probability that at least one UD experiences a deep fade, which limits the max-min objective. Nevertheless, the proposed PD-TQC achieves the highest minimum uplink throughput for all values of $\kappa_R$, with average improvements of $78.08\%$, $15.90\%$, $3.50\%$, and $2.53\%$ over the Greedy, PPO, SAC, and PD-TQC with SDMA-based SWIPT baselines, respectively, while satisfying the downlink throughput and UAV residual-energy constraints in at least $96.99\%$ and $98.60\%$ of the evaluated frames.

Moreover, the benefit of RSMA becomes more pronounced as $\kappa_R$ increases. Under a large $\kappa_R$, the LoS-dominated UD channels become highly correlated, causing strong inter-stream interference that RSMA mitigates via the common stream. Accordingly, the average common-stream power ratio $\beta_c$ of the learned PD-TQC policy increases from $0.01$ at $\kappa_R=-5$~dB to $0.66$ at $\kappa_R=10$~dB, whereas the per-frame downlink throughput constraint satisfaction rate of the SDMA-based scheme decreases from $99.07\%$ to $85.91\%$. As $\kappa_R$ decreases, the UD channels become less correlated, and RSMA adaptively approaches its special case SDMA by reducing $\beta_c$, so that the two schemes achieve similar performance under scattering-dominated conditions.

\subsubsection{Impact of Number of UDs}
\begin{figure}[!t]
\centering
\includegraphics[width=3.4in]{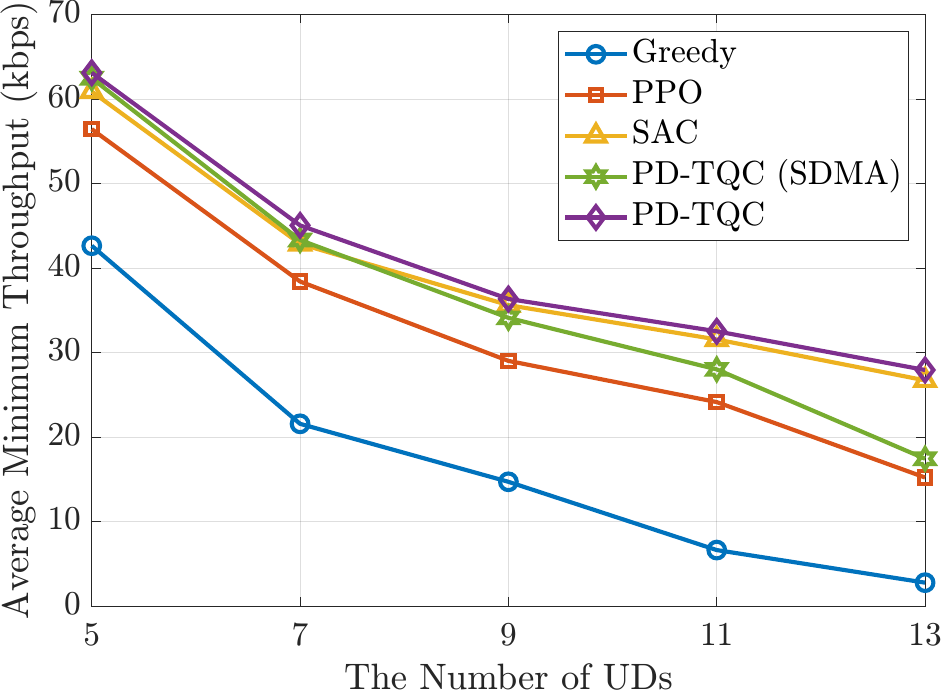}
\caption{Annual average minimum uplink throughput versus the number of UDs for the UAV-enabled WPUSN system using the 2025 annual VWC dataset, evaluated with the trained policies of the Greedy, PPO, SAC, and proposed PD-TQC schemes, as well as the PD-TQC scheme with SDMA-based SWIPT.}
\label{fig_nnres}
\vspace{-2mm}
\end{figure}

Fig.~\ref{fig_nnres} depicts the annual average minimum uplink throughput under varying numbers of UDs, i.e., $K\in\{5,7,9,11,13\}$, for the considered UAV-enabled WPUSN system using the 2025 annual VWC dataset, evaluated using the trained policies of the Greedy, PPO, SAC, PD-TQC with SDMA-based SWIPT, and the proposed PD-TQC schemes. One can see that the annual average minimum uplink throughput decreases for all approaches as the number of UDs increases from $5$ to $13$ due to increased competition for limited resources and the higher probability of UDs experiencing poor channel conditions. Nevertheless, the proposed PD-TQC consistently achieves the highest minimum uplink throughput across different numbers of UDs, delivering average gains of $131.98\%$, $25.56\%$, $3.71\%$, and $10.58\%$ over the Greedy, PPO, SAC, and PD-TQC with SDMA-based SWIPT baselines, respectively, while satisfying the downlink throughput and UAV residual-energy constraints in at least $96.99\%$ and $99.20\%$ of the evaluated frames. Moreover, as $K$ increases from $5$ to $13$, the state and action dimensions of PD-TQC increase linearly from $23$ to $55$, whereas the online computation increases only from $0.30$ to $0.35$ million floating-point operations per decision, with an inference latency of at most $1.07$~ms. These results demonstrate the superior capability of the proposed PD-TQC framework in preserving uplink fairness as the network scales.

\subsubsection{Impact of Downlink Throughput Threshold}
\begin{figure}[!t]
\centering
\includegraphics[width=3.4in]{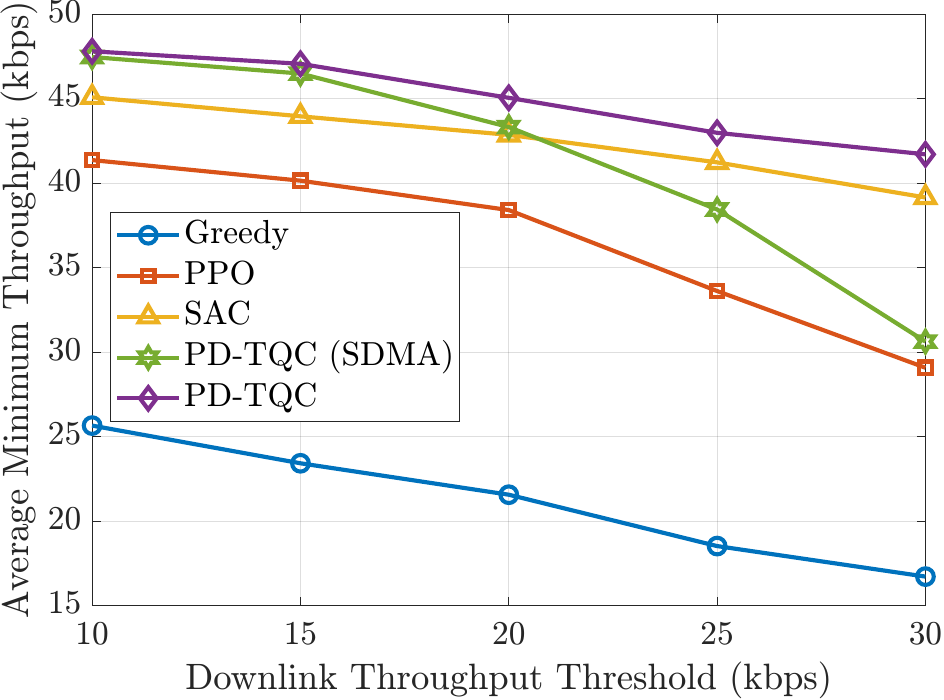}
\caption{Annual average minimum uplink throughput versus the downlink throughput threshold for the UAV-enabled WPUSN system using the 2025 annual VWC dataset, evaluated with the trained policies of the Greedy, PPO, SAC, and proposed PD-TQC schemes, as well as the PD-TQC scheme with SDMA-based SWIPT.}
\label{fig_ttres}
\end{figure}

Fig.~\ref{fig_ttres} illustrates the annual average minimum uplink throughput versus the downlink throughput threshold for the considered UAV-enabled WPUSN system using the 2025 annual VWC dataset, evaluated with the trained policies of the Greedy, PPO, SAC, PD-TQC with SDMA-based SWIPT, and the proposed PD-TQC schemes. As the downlink throughput threshold increases from $10$~kbps to $30$~kbps, the annual average minimum uplink throughput of all schemes decreases since a more stringent downlink constraint forces the agent to allocate more resources to satisfy the required downlink throughput, thereby leaving fewer degrees of freedom for uplink fairness optimization. Across all threshold settings, the proposed PD-TQC consistently achieves the highest minimum uplink throughput, providing average improvements of $112.02\%$, $22.98\%$, $5.80\%$, and $8.85\%$ over the Greedy, PPO, SAC, and PD-TQC with SDMA-based SWIPT baselines, respectively. Moreover, when $R_{\min}=30$~kbps, the annual average minimum downlink throughput of the SDMA-based PD-TQC drops to only $29.72$~kbps, i.e., below the threshold, whereas the proposed PD-TQC attains $46.74$~kbps while sustaining an annual average minimum uplink throughput of $41.71$~kbps. This gain stems from RSMA, which enables more flexible interference management and thus enlarges the feasible region under stringent downlink throughput requirements. Overall, these results show that the proposed PD-TQC approach can balance uplink fairness maximization and downlink throughput enforcement under time-varying soil-air channels.

\section{Conclusion} \label{sec:conclusion}
This paper extended the UAV-enabled WPUSN architecture by integrating an RSMA-based SWIPT downlink with a TDMA-based uplink, thereby enabling a more complete and sustainable IoUT framework. Within this system, we investigated the max-min uplink throughput problem under per-UD downlink throughput requirements and UAV energy-consumption constraints. Considering time-varying soil-air channels, the optimization problem was first reformulated as a constrained MDP with a per-frame max-min reward and violation-based constraint costs. Then, we proposed the PD-TQC solution, where a TQC-based reward critic evaluated the throughput fairness objective, two pairs of twin cost critics learned the two constraint costs, and a stochastic actor generated the resource-control policy with adaptive dual-variable updates. Through simulations in a realistic underground monitoring case based on annual \textit{in-situ} soil VWC measurements, our numerical results demonstrated that the proposed PD-TQC approach consistently outperformed the four benchmark schemes in terms of annual average minimum uplink throughput across different Rician factors, numbers of UDs, and downlink throughput thresholds. These results provided a basis for future experimental validation and further investigation of DRL-enabled resource allocation in UAV-assisted IoUT systems. Note that these results were obtained at a single irrigation site with fixed soil texture, burial depth, and deployment parameters, and thus their generalization to other soil textures and climates remains to be verified. Nevertheless, the proposed RS-SWIPT framework can be adapted to other underground monitoring scenarios through scenario-specific channel modeling, parameter calibration, and agent retraining. Its current implementation is primarily suited to shallow underground monitoring, as severe soil-induced attenuation at greater burial depths compromises RS-SWIPT performance. Future work will therefore investigate the integration of the proposed framework with the Terra-Fi architecture developed in~\cite{LinTerraFi2026} to support deep underground monitoring while maintaining reliable wireless communication and efficient WET operation.

\bibliographystyle{IEEEtran}
\bibliography{ref}

\end{document}